\documentclass[longauth]{aa}
\usepackage{booktabs}
\usepackage{array}
\usepackage{bm}
\usepackage{CJK}
\usepackage{etoolbox} \usepackage[slantedGreek]{newtxmath}
\usepackage{subfigure}
\usepackage{graphicx}
\usepackage{color}
\usepackage{siunitx}
\usepackage{multirow}
\usepackage{ulem}
\usepackage{xspace}
\usepackage[svgnames]{xcolor}

\newcommand{\cntext}[1]{\begin{CJK}{UTF8}{gbsn}#1\end{CJK}}
\DeclareSIUnit{\jansky}{Jy}
\newcommand{\pkstwo}{PKS~2131$-$021\xspace}
\newcommand{\pkszero}{PKS~J0805$-$0111\xspace}
\newcommand{\snrsin}{\ensuremath{\mathrm{SNR}_{\mathrm{sin}}}}
\newcommand{\snrvar}{\ensuremath{\mathrm{SNR}_{\mathrm{var}}}}

\begin{document} 

\title{The Atacama Cosmology Telescope: Observations\\of supermassive black hole binary candidates}
\subtitle{Strong sinusoidal variations at 95, 147 and 225\,GHz\\in \pkstwo and \pkszero}
\titlerunning{ACT Observations of Sinusoidal Variations in SMBHB Candidates}
\authorrunning{A.~D.~Hincks et al.}

\author{Adam~D.~Hincks\inst{1,2} \and Xiaoyi~Ma~(\cntext{马潇依})\inst{3,4} \and Przemek~Mr{\'o}z\inst{5} \and Sigurd~K.~Naess\inst{6} \and Sebastian~Kiehlmann\inst{7} \and Roger~D.~Blandford\inst{8} \and J.~Richard~Bond\inst{9} \and Mark~Devlin\inst{10} \and Jo~Dunkley\inst{11,12} \and Allen~Foster\inst{11} \and Matthew~J.~Graham\inst{13} \and Yilun~Guan\inst{14} \and Carlos~Herv{\'i}as-Caimapo\inst{15} \and John~C.~Hood~II\inst{16} \and Arthur~Kosowsky\inst{17} \and Aretaios~Lalakos\inst{18,19} \and Elias~R.~Most\inst{18,19} \and Michael~D.~Niemack\inst{20,21} \and John~Orlowski-Scherer\inst{10} \and Lyman~A.~Page\inst{11} \and Bruce~Partridge\inst{22} \and Anthony~C.~S.~Readhead\inst{23} \and Crist\'obal~Sif\'on\inst{24} \and Suzanne~T.~Staggs\inst{11} \and Andrew~G.~Sullivan\inst{8} \and Cristian~Vargas\inst{25}}
\institute{
  {David A. Dunlap Department of Astronomy \& Astrophysics, University of Toronto, 50 St. George St., Toronto ON M5S 3H4, Canada} \and
  {Specola Vaticana (Vatican Observatory), V-00120 Vatican City State} \and
  {Kavli Institute for Astronomy \& Astrophysics, Peking University, Beijing 100871, People's Republic of China} \and
  {Department of Astronomy, School of Physics, Peking University, Beijing 100871, People's Republic of China} \and
  {Astronomical Observatory, University of Warsaw, Al. Ujazdowskie 4, 00-478 Warszawa, Poland} \and
  {Institute of Theoretical Astrophysics, University of Oslo, Norway} \and
  {Institute of Astrophysics, Foundation for Research \& Technology-Hellas, GR-71110 Heraklion, Greece} \and
  {Kavli Institute for Particle Astrophysics and Cosmology, Department of Physics, Stanford University, Stanford, CA 94305, USA} \and
  {Canadian Institute for Theoretical Astrophysics, University of Toronto, 60 St. George Street, Toronto, ON M5S 3H8, Canada} \and
  {Department of Physics \& Astronomy, University of Pennsylvania, 209 South 33rd Street, Philadelphia, PA, 19104, USA} \and
  {Joseph Henry Laboratories of Physics, Jadwin Hall, Princeton University, Princeton, NJ 08544, USA} \and
  {Department of Astrophysical Sciences, Peyton Hall, Princeton University, Princeton, NJ 08544, USA} \and
  {Division of Physics, Mathematics, and Astronomy, California Institute of Technology, Pasadena, CA 91125, USA} \and
  {Dunlap Institute for Astronomy \& Astrophysics, University of Toronto, 50 St. George St., Toronto ON M5S 3H4, Canada} \and
  {Instituto de Astrof\'isica \& Centro de Astro-Ingenier\'ia, Facultad de F\'isica, Pontificia Universidad Cat\'olica de Chile, Av. Vicu\~na Mackenna 4860, 7820436 Macul, Santiago, Chile} \and
  {Department of Astronomy \& Astrophysics, University of Chicago, 5640 South Ellis Avenue, Chicago, IL 60637, USA} \and
  {Department of Physics and Astronomy, University of Pittsburgh, 3941 O'Hara Street, Pittsburgh, PA 15260, USA} \and
  {TAPIR, Mailcode 350-17, California Institute of Technology, Pasadena, CA 91125, USA} \and
  {Walter Burke Institute for Theoretical Physics, California Institute of Technology, Pasadena, CA 91125, USA} \and
  {Department of Physics, Cornell University, Ithaca, NY 14853, USA} \and
  {Department of Astronomy, Cornell University, Ithaca, NY 14853, USA} \and
  {Department of Physics \& Astronomy, Haverford College, Haverford, PA 19041, USA} \and
  {Owens Valley Radio Observatory, California Institute of Technology, Pasadena, CA 91125, USA} \and
  {Instituto de F\'isica, Pontificia Universidad Cat\'olica de Valpara\'iso, Casilla 4059, Valpara\'iso, Chile} \and
  {Mitchell Institute for Fundamental Physics \& Astronomy and Department of Physics \& Astronomy, Texas A\&M University, College Station, TX 77843, USA}
}
 
\abstract{
Large sinusoidal variations in the radio light curves of the blazars \pkszero and \pkstwo have recently been discovered with an 18-year monitoring programme at the Owens Valley Radio Observatory, making these systems strong supermassive black hole binary (SMBHB) candidates. The sinusoidal variations in \pkstwo dominate its light curves from 2.7\,GHz to optical frequencies. We report sinusoidal variations observed in both objects with the Atacama Cosmology Telescope (ACT) at 95, 147 and 225\,GHz consistent with the radio light curves. The ACT 95\,GHz light curve of \pkstwo agrees well with the contemporaneous 91.5\,GHz ALMA light curve and is comparable in quality, while the ACT light curves of \pkszero, for which there are no ALMA or other millimetre light curves, show that \pkstwo is not an isolated case, and that this class of AGN exhibits the following properties: (a) the sinusoidal pattern dominates over a broad range of frequencies; (b) the amplitude of the sine wave compared to its mean value is monochromatic (i.e., nearly constant across frequencies); (c) the phase of the sinusoid phase changes monotonically as a function of frequency; (d) the sinusoidal variations are intermittent. We describe a physical model for SMBHB systems, the modified Kinetic Orbital model, that explains all four of these phenomena. Monitoring of ${\sim}8000$ blazars by the Simons Observatory over the next decade should provide a large number of SMBHB candidates that will shed light on the nature of the nanohertz gravitational-wave background.
}
\maketitle
\section{Introduction}\label{sec:intro}

Evidence of a stochastic background of gravitational waves with periods of months to years has recently been presented by the North American Nanohertz Observatory for Gravitational Waves collaboration \citep{agazie/etal:2023a} and the European Pulsar Timing Array collaboration \citep{epta:2023}; the MeerKAT Pulsar Timing Array collaboration also find evidence for the background, but caution that it is highly dependent on choices in their noise modelling \citep{miles/etal:2025}. These experiments use millisecond pulsar timing arrays 
(\citealt{1974MmRAS..78....1R,1982Natur.300..615B})\footnote{There were two crucial steps in the discovery of millisecond pulsars: (i) the discovery of interplanetary scintillation, at Galactic latitude $-0.3^\circ$, in 4C~21.53 by \citet{1974MmRAS..78....1R} (see \citealt{2024JAHH...27..453R}), which drew attention to the singular nature of this object; and (ii) the discovery of millisecond pulses from 4C~21.53W by \citet{1982Natur.300..615B}.} which currently provide the most sensitive method of searching for gravitational waves at these frequencies, although stellar astrometry of billions of stars \citep[e.g.,][]{2017PhRvL.119z1102M,1990NCimB.105.1141B} may be competitive in the future. 
Supermassive black hole binary systems (SMBHBs) have been suggested as the origin of this stochastic background \citep[e.g.,][]{agazie/etal:2023a,agazie/etal:2023b}, making clear the importance of searches for their electromagnetic counterparts.

Supermassive black holes (SMBHs) are the central engines in active galactic nuclei (AGN), which are strong sources of emission across the electromagnetic spectrum \citep{2000AmJPh..68..489K,Blandford2019}. Since 2008  the 40\,m Telescope of the Owens Valley Radio Observatory (OVRO) has been dedicated full time to monitoring ${\sim}1830$ blazars at 15 GHz on a 3--4 day cadence \citep{Richards2011}. 
Of the  ${\sim}1830$ blazars monitored, 1158 comprise a complete sample suitable for statistical tests. 
To date, two AGN with sinusoidal variations dominating their light curves have been identified in this sample, which rigorous statistical tests show are unlikely to be produced by random fluctuations in the red noise tail of their power spectral density: \pkstwo (\citealt{2022ApJ...926L..35O}, hereafter referred to as O22; \citealt{2025ApJ...985...59K}, hereafter K25) and \pkszero (\citealt{delaparra/etal:2025}, hereafter 
D25). K25 and D25 simulated $10^6$ light curves that match the power spectral densities and probability density functions (i.e., the flux distribution) of \pkstwo and \pkszero, and showed that the joint global probability, $p$, of these two sources, out of the sample of 1158, being generated randomly due to the red noise tail in their variability spectra is $p<3\times 10^{-3}$ (D25).
\citet{ren/etal:2021b} and \citet{ren/etal:2021} were the first to draw attention to possible quasi-periodic oscillations in \pkszero and \pkstwo, respectively. However, in the case of \pkszero, D25 argue that they overestimated the significance of the periodicity by several orders of magnitude by assuming the data consisted of independent Gaussian values; similarly, for \pkstwo, the reported significance of the periodicity is inflated, since they consider the local rather than global significance (c.f., O22).

It is by no means obvious that orbital motion of a SMBH with a relativistic jet will produce sinusoidal variations; however, a mechanism was discovered independently by \citet{2017MNRAS.465..161S} and by O22, which K25 refer to as the `Kinetic Orbital' (KO) model. In the KO model, one of the SMBHs in the binary system produces a jet, and aberration of this jet due to orbital motion has a large effect on the observed emission from the highly relativistic emitting material. This is shown in O22 to produce sinusoidal variations. While other mechanisms like oscillations or precessions in the accretion disc have been invoked to explain (quasi-)periodic behaviour in AGN light curves (see, e.g., discussions in O22; \citealt{TaoAn2013,wangpvb2014,Ingramreview}), the KO model is remarkably simple and economically explains the relevant observed phenomena.

In addition to studying radio light curves, K25 analysed publicly available light curves of \pkstwo at millimetre (mm), infrared and optical frequencies and found the same sine wave pattern in them all, as well as `hints' of a sinusoid at $\gamma$-ray frequencies. This broad-band signal is explained well by the KO model since the orbital aberration affects all frequencies. The light curves exhibited a monotonic phase shift in the sine waves from the radio to the optical, spanning more than five decades in frequency. This can be interpreted as an optical depth effect in which the higher frequencies probe closer to the central engine, a behaviour that has been observed in many blazars since it was discovered in the first maps made with very long baseline interferometry \citep{1978Natur.276..768R,1980IAUS...92..165R}. In the KO model, simultaneous observations of multiple frequencies capture the jet at different points along its helical structure, and thus at different phases of the orbital pattern. Although the light travel time between the higher-frequency and lower-frequency zones corresponds to several sinusoidal cycles, the jet itself is traveling at close to the speed of light. Therefore, by the time the higher-frequency light arrives at the location of the lower-frequency emission, the jet material lags behind only slightly, leading to a relatively small phase delay in the observed light curve. An issue with this model, however, is that it is unclear how the helical structure can remain coherent over a large number of windings spanning tens of parsecs.

In this paper we report observations with the Atacama Cosmology Telescope (ACT) of \pkstwo and \pkszero from 2016 to 2022 at 95, 147 and 225\,GHz that provide new insights into these two objects and the phenomenology of SMBHBs in blazars. Furthermore, we present a modified KO (MKO) model that solves the problem of maintaining a helical structure of the jet over many orbital periods by proposing that a sub-relativistic wind, which is dragged by the SMBH orbit, confines the relativistic flow of the jet and results in a helix that need not span so many windings to explain the observations. We also review recent numerical simulations of accretion discs that show how the jet can be interrupted in such a way as to produce the observed intermittency of the sinusoidal pattern in the light curves. Finally, we discuss the potential for wide-field CMB surveys to discover scores of SMBHB candidates and determine their periods, which has implications for gravitational wave science and SMBHB population studies.

The paper is organised as follows. In Sec.~\ref{sec:data} we describe the ACT light curves and briefly introduce other key data used in our analyses. In Sec.~\ref{sec:sinfit}, we present sinusoidal fits to the ACT light curves of \pkstwo and \pkszero, and in Sec.~\ref{sec:multifreq_properties}, we analyse and discuss the relative phase shifts between light curves at different frequencies as well as the achromaticity of the sinusoidal amplitudes. In Sec.~\ref{sec:newmodsims}, we present the new, MKO model and outline the relevance of recent numerical simulations of accretion discs for SMBHB blazars. In Sec.~\ref{sec:cmparison}, we discuss the promise of mm observations of AGN for future SMBHB candidate discoveries, comparing and contrasting to optical searches. We summarise our findings and discuss the potential of upcoming mm-surveys for discovering SMBHB candidates in Sec.~\ref{sec:discussion}.

\section{Data}\label{sec:data}

\subsection{The ACT observations}\label{sec:actobs}

ACT was a cosmic microwave background (CMB) experiment that operated from 2007 to 2022 in the Atacama Desert of Chile. Its 6\,m telescope observed with three generations of receivers: the Millimeter Bolometric Array Camera (MBAC, 2007--11; \citealt{swetz/etal:2011}), ACTPol (2013\mbox{--}15; \citealt{thornton/etal:2016}) and Advanced ACTPol (2016--22; \citealt{henderson/etal:2016}). All receivers were equipped with three optics tubes, each terminating in its own array of detectors that were occasionally changed to allow different combinations of frequencies to be observed. Starting with {ACTPol}, detectors were polarisation-sensitive and, beginning with the fourth polarised array (PA4), were also dichroic, i.e., sensitive to two frequency bands. Collectively, PA1 to PA7 observed in five frequency bands: f030, f040, f090, f150 and f220. The two lowest frequencies are still being analysed, and all science results so far, including in this paper, come from f090, f150 and f220, whose band centres for a synchrotron spectral index of $-0.7$, averaged across the PAs, were 95.0, 147.1 and 225.0\,GHz, respectively.\footnote{For f150 we have given the mean excluding PA1, since that array was removed before the light curves in this paper begin. With PA1 included the mean band centre of f150 is 146.9\,GHz.} See \citet{hervias-caimapo/etal:2024} for a full summary of the array timelines and frequencies. The  passbands are discussed in \citet{madhavacheril/etal:2020} and \citet{coulton/etal:2024}.

A paper presenting the ACT bright AGN sample is in preparation (Ma et al.), and the light curve data will be publicly released. The data reduction process and an analysis of calibration uncertainties and other error sources will be described more fully in that paper and here we introduce only the essentials. Light curves are generated by creating $2\times2\,\si{deg\squared}$ maps centred on the source for each day the source was observed. Data are calibrated from raw readout units to incident power by measuring the detector response to small, square-wave changes in the voltage bias applied to the detectors \citep{niemack:2008}; this is performed at least hourly. Interdetector calibration is achieved by flat fielding on the atmospheric emission which creates a common signal mode across the detectors. Calibration from incident power to celestial flux density is achieved using observations of Uranus that were taken every few nights and a final correction based on cross correlating ACT and \textit{Planck} angular power spectra is applied.\footnote{The 147\,GHz data from PA4 are lacking the final, \textit{Planck}-based correction due to issues with the ACT power spectra for these data \citep{naess/etal:2025}. Since these corrections for other frequency-arrays are on the order of a few percent, comparable to our uncertainty (see below), this should not noticeably affect our results in this paper.} See \citet{choi/etal:2020} and \citet{naess/etal:2025} for more information on the general map-making process. With the individual maps in hand, the point source flux density is extracted from each with a matched filter using the known telescope beams and a noise covariance estimated from the map itself after masking the source, similar to the method of \citet{marriage/etal:2011}, except that individual pixel weights are also accounted for in the noise covariance.\footnote{For details, see the documentation for the \texttt{matched\_filter\_constcorr\_dual()} method in the \texttt{pixell} package: \url{https://github.com/simonsobs/pixell}.} The resulting light curves tend to have a few outlying points, almost all of which are automatically flagged as outliers by our pipeline due to the low effective number of detectors in the receiver for a particular observation (e.g., because of bad weather, or because the source was near the edge of the observed field). A small fraction of data points, ${\sim}0.1\%$, are flagged as outliers based on conservative visual inspections. In this work we exclude all points flagged as outliers and we combine all light curves from different polarised arrays that are observed at the same frequency (e.g., PA3--f090, PA5--f090 and PA6--f090).

Ma et al. will provide a detailed uncertainty analysis of the calibration, but preliminary studies of the variance of Uranus measurements over time, as well as comparison of flux densities of the AGN between array-bands (e.g., PA5--f090 vs. PA6--f090), indicate an uncertainty of ${\sim}2$ to 7\%, depending on the array-frequency channel \citep[c.f.,][]{hervias-caimapo/etal:2024}. These systematic measurement errors, which are not Gaussian-distributed, cause point-to-point scatter in the light curves that is mixed in with any intrinsic variability of observed sources. However, the results of this paper do not rely on detailed knowledge of these sources of uncertainty.

\subsection{Key Data at Other Wavelengths}\label{sec:other_data}

Our analysis includes the following light curves from other observatories that overlap in time with the ACT light curves:
\begin{itemize}
    \item OVRO light curves of \pkstwo and \pkszero at 15\,GHz \citep{Richards2011}.
    \item Atacama Large Millimeter Array (ALMA) light curves of \pkstwo at 91.5, 103, 104, 337, 343, 344, 349 and 350\,GHz, taken from the ALMA Calibrator Source Catalogue.\footnote{\url{https://almascience.eso.org/alma-data/calibrator-catalogue}} Following K25, we combine the 103 and 104\,GHz data into a single light curve and call the combination `103.5\,GHz'. We also combine 337--350\,GHz and refer to it as `345\,GHz'.
    \item The Catalina Real Time Survey (CRTS; \citealt{drake2009}) light curve of \pkstwo in the optical $V$-band.
    \item The Zwicky Transient Facility (ZTF; \citealt{masci2019,graham2019}) light curve of \pkstwo in the optical $g$-band.
\end{itemize}
Data from CRTS and ZTF were rebinned to a daily cadence using a weighted mean scheme. 
For consistency with K25, especially in our analysis of the sinusoid phase shifts (Sec.~\ref{sec:phase_shifts}), we use the same time spans as they did: MJD 56647.1--60053.7 (21 December 2013 -- 19 April 2023) for the ALMA light curves, and MJD 54470.9--60175.3 (5 January 2008 -- 19 August 2023) for OVRO.

The redshifts of \pkstwo and \pkszero are 1.285 and 1.39, respectively \citep{drinkwater/etal:1997,paliya/etal:2017}.
 
\begin{figure*}[t]
   \centering
   \footnotesize
\begingroup
  \makeatletter
  \providecommand\color[2][]{\GenericError{(gnuplot) \space\space\space\@spaces}{Package color not loaded in conjunction with
      terminal option `colourtext'}{See the gnuplot documentation for explanation.}{Either use 'blacktext' in gnuplot or load the package
      color.sty in LaTeX.}\renewcommand\color[2][]{}}\providecommand\includegraphics[2][]{\GenericError{(gnuplot) \space\space\space\@spaces}{Package graphicx or graphics not loaded}{See the gnuplot documentation for explanation.}{The gnuplot epslatex terminal needs graphicx.sty or graphics.sty.}\renewcommand\includegraphics[2][]{}}\providecommand\rotatebox[2]{#2}\@ifundefined{ifGPcolor}{\newif\ifGPcolor
    \GPcolortrue
  }{}\@ifundefined{ifGPblacktext}{\newif\ifGPblacktext
    \GPblacktexttrue
  }{}\let\gplgaddtomacro\g@addto@macro
\gdef\gplbacktext{}\gdef\gplfronttext{}\makeatother
  \ifGPblacktext
\def\colorrgb#1{}\def\colorgray#1{}\else
\ifGPcolor
      \def\colorrgb#1{\color[rgb]{#1}}\def\colorgray#1{\color[gray]{#1}}\expandafter\def\csname LTw\endcsname{\color{white}}\expandafter\def\csname LTb\endcsname{\color{black}}\expandafter\def\csname LTa\endcsname{\color{black}}\expandafter\def\csname LT0\endcsname{\color[rgb]{1,0,0}}\expandafter\def\csname LT1\endcsname{\color[rgb]{0,1,0}}\expandafter\def\csname LT2\endcsname{\color[rgb]{0,0,1}}\expandafter\def\csname LT3\endcsname{\color[rgb]{1,0,1}}\expandafter\def\csname LT4\endcsname{\color[rgb]{0,1,1}}\expandafter\def\csname LT5\endcsname{\color[rgb]{1,1,0}}\expandafter\def\csname LT6\endcsname{\color[rgb]{0,0,0}}\expandafter\def\csname LT7\endcsname{\color[rgb]{1,0.3,0}}\expandafter\def\csname LT8\endcsname{\color[rgb]{0.5,0.5,0.5}}\else
\def\colorrgb#1{\color{black}}\def\colorgray#1{\color[gray]{#1}}\expandafter\def\csname LTw\endcsname{\color{white}}\expandafter\def\csname LTb\endcsname{\color{black}}\expandafter\def\csname LTa\endcsname{\color{black}}\expandafter\def\csname LT0\endcsname{\color{black}}\expandafter\def\csname LT1\endcsname{\color{black}}\expandafter\def\csname LT2\endcsname{\color{black}}\expandafter\def\csname LT3\endcsname{\color{black}}\expandafter\def\csname LT4\endcsname{\color{black}}\expandafter\def\csname LT5\endcsname{\color{black}}\expandafter\def\csname LT6\endcsname{\color{black}}\expandafter\def\csname LT7\endcsname{\color{black}}\expandafter\def\csname LT8\endcsname{\color{black}}\fi
  \fi
    \setlength{\unitlength}{0.0500bp}\ifx\gptboxheight\undefined \newlength{\gptboxheight}\newlength{\gptboxwidth}\newsavebox{\gptboxtext}\fi \setlength{\fboxrule}{0.5pt}\setlength{\fboxsep}{1pt}\definecolor{tbcol}{rgb}{1,1,1}\begin{picture}(10080.00,5320.00)\gplgaddtomacro\gplbacktext{\csname LTb\endcsname \put(685,530){\makebox(0,0)[r]{\strut{}$0$}}\csname LTb\endcsname \put(685,1217){\makebox(0,0)[r]{\strut{}$0.5$}}\csname LTb\endcsname \put(685,1904){\makebox(0,0)[r]{\strut{}$1$}}\csname LTb\endcsname \put(685,2590){\makebox(0,0)[r]{\strut{}$1.5$}}\csname LTb\endcsname \put(685,3277){\makebox(0,0)[r]{\strut{}$2$}}\csname LTb\endcsname \put(685,3964){\makebox(0,0)[r]{\strut{}$2.5$}}\csname LTb\endcsname \put(1741,393){\makebox(0,0){\strut{}$55000$}}\csname LTb\endcsname \put(3339,393){\makebox(0,0){\strut{}$56000$}}\csname LTb\endcsname \put(4937,393){\makebox(0,0){\strut{}$57000$}}\csname LTb\endcsname \put(6535,393){\makebox(0,0){\strut{}$58000$}}\csname LTb\endcsname \put(8132,393){\makebox(0,0){\strut{}$59000$}}\csname LTb\endcsname \put(9730,393){\makebox(0,0){\strut{}$60000$}}\csname LTb\endcsname \put(888,4651){\makebox(0,0){\strut{}2008}}\csname LTb\endcsname \put(1473,4651){\makebox(0,0){\strut{}2009}}\csname LTb\endcsname \put(2056,4651){\makebox(0,0){\strut{}2010}}\csname LTb\endcsname \put(2639,4651){\makebox(0,0){\strut{}2011}}\csname LTb\endcsname \put(3223,4651){\makebox(0,0){\strut{}2012}}\csname LTb\endcsname \put(3807,4651){\makebox(0,0){\strut{}2013}}\csname LTb\endcsname \put(4390,4651){\makebox(0,0){\strut{}2014}}\csname LTb\endcsname \put(4974,4651){\makebox(0,0){\strut{}2015}}\csname LTb\endcsname \put(5557,4651){\makebox(0,0){\strut{}2016}}\csname LTb\endcsname \put(6142,4651){\makebox(0,0){\strut{}2017}}\csname LTb\endcsname \put(6725,4651){\makebox(0,0){\strut{}2018}}\csname LTb\endcsname \put(7308,4651){\makebox(0,0){\strut{}2019}}\csname LTb\endcsname \put(7891,4651){\makebox(0,0){\strut{}2020}}\csname LTb\endcsname \put(8476,4651){\makebox(0,0){\strut{}2021}}\csname LTb\endcsname \put(9059,4651){\makebox(0,0){\strut{}2022}}\csname LTb\endcsname \put(9642,4651){\makebox(0,0){\strut{}2023}}}\gplgaddtomacro\gplfronttext{\csname LTb\endcsname \put(1547,1690){\makebox(0,0)[l]{\strut{}OVRO 15\,GHz}}\csname LTb\endcsname \put(1547,1494){\makebox(0,0)[l]{\strut{}ALMA 91.5\,GHz}}\csname LTb\endcsname \put(1547,1297){\makebox(0,0)[l]{\strut{}ACT 95\,GHz}}\csname LTb\endcsname \put(1547,1100){\makebox(0,0)[l]{\strut{}ACT 147\,GHz}}\csname LTb\endcsname \put(1547,904){\makebox(0,0)[l]{\strut{}ACT 225\,GHz}}\csname LTb\endcsname \put(1547,707){\makebox(0,0)[l]{\strut{}ALMA 345\,GHz}}\csname LTb\endcsname \put(219,2522){\rotatebox{-270.00}{\makebox(0,0){\strut{}Flux Density (Jy)}}}\csname LTb\endcsname \put(5416,137){\makebox(0,0){\strut{}Modified Julian Day}}\csname LTb\endcsname \put(5416,5005){\makebox(0,0){\strut{}\textbf{\pkstwo}}}}\gplbacktext
    \put(0,0){\includegraphics[width={504.00bp},height={266.00bp}]{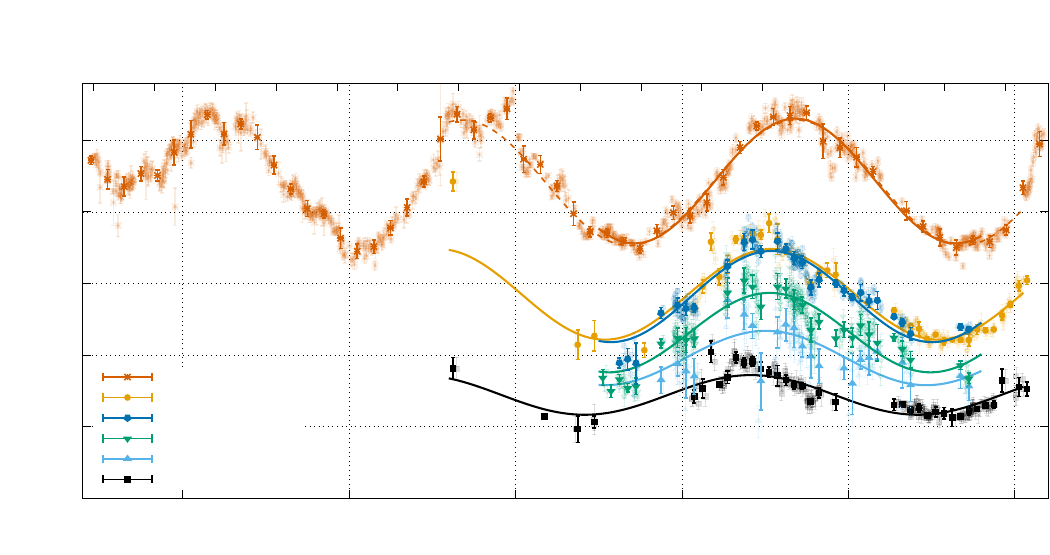}}\gplfronttext
  \end{picture}\endgroup
       \caption{\pkstwo light curves from OVRO, ACT and ALMA. Heavy points are binned into 50-day intervals (100 for OVRO) to guide the eye, with individual measurements shown as lighter points. Sinusoidal fits, shown with the continuous curves, were done on the unbinned data. The sine fits for ALMA are from K25 and sine fits to ACT are done in this paper (see text). For the OVRO data, the solid curve is the best-fit for the ACT time range, while the dashed sine curve corresponds to the ALMA time range. A monotonic phase shift of the sine waves to earlier times with increasing frequency is visible by eye, and is reported relative to the OVRO phase as $\Delta\phi_0$ in Tables~\ref{tab:2131_fits} and \ref{tab:alma_2131_fits}.}
         \label{plt:2131}
\end{figure*}

\begin{table*}[h]
\centering
\caption{Sine-wave fit results for \pkstwo.}
\begin{tabular}{lrrrr}
\hline \hline
Parameter & OVRO 15\,GHz & ACT 95\,GHz & ACT 147\,GHz & ACT\,225 GHz \\
\hline
$P$ (days)              & \multicolumn{4}{c}{$1931\pm19$ (fixed)} \\
$A$   (Jy)              & $0.4357 \pm 0.0052$ & $0.3185 \pm 0.0089$  & $0.2776 \pm 0.0078$  & $0.1899 \pm 0.0240$ \\
$\phi_0$ (rad)          & $3.678 \pm 0.015$   & $3.212 \pm 0.025$    & $3.179 \pm 0.027$    & $3.100 \pm 0.111$ \\
$\bar{S}$ (Jy)          & $2.2148 \pm 0.0040$ & $1.4070 \pm 0.0061$  & $1.1550 \pm 0.0054$  & $0.9773 \pm 0.0165$ \\
$\xi$ (Jy)              & $0.0656 \pm 0.0032$ & $0.0897 \pm 0.0043$  & $0.0992 \pm 0.0040$  & $0.1374 \pm 0.0141$ \\
$\Delta\phi_0$ (cycles) & 0                   & $-0.0742 \pm 0.0046$ & $-0.0793 \pm 0.0048$ & $-0.0919 \pm 0.0179$ \\
$\Delta\phi_0$ (days)   & 0                   & $-143.2 \pm 9.0$     & $-153.1 \pm 9.3$     & $-177.5 \pm 34.6$ \\
\hline
\end{tabular}
\tablefoot{The period, $P$, was determined from an initial fit including only the OVRO data in the range $57500 < \mathrm{MJD} < 59800$. The joint fit including OVRO and the three ACT light curves kept the period fixed at this best fitting value. The fit uncertainties do not account for systematics due to superimposed, shorter term variations (see Appendix~\ref{app:empirical_estimate}). The reference time for the phase, in MJD, is  $t_0 = 59\,000$. The phase shift, $\Delta\phi_0$, is defined in Eq.~\ref{eq:phase_shift}. Here, and in Tables~\ref{tab:0805_fits} and \ref{tab:alma_2131_fits}, two significant figures are retained in the quoted uncertainties for consistency across all fits and as useful for error propagation, but we do not claim to know errors to that precision (see Appendix~\ref{app:empirical_estimate_backgnd}).}
\label{tab:2131_fits}
\end{table*}

\section{Sine-wave fits to the light curves}\label{sec:sinfit}

\begin{figure}[t]
   \centering
   \begingroup
  \makeatletter
  \providecommand\color[2][]{\GenericError{(gnuplot) \space\space\space\@spaces}{Package color not loaded in conjunction with
      terminal option `colourtext'}{See the gnuplot documentation for explanation.}{Either use 'blacktext' in gnuplot or load the package
      color.sty in LaTeX.}\renewcommand\color[2][]{}}\providecommand\includegraphics[2][]{\GenericError{(gnuplot) \space\space\space\@spaces}{Package graphicx or graphics not loaded}{See the gnuplot documentation for explanation.}{The gnuplot epslatex terminal needs graphicx.sty or graphics.sty.}\renewcommand\includegraphics[2][]{}}\providecommand\rotatebox[2]{#2}\@ifundefined{ifGPcolor}{\newif\ifGPcolor
    \GPcolortrue
  }{}\@ifundefined{ifGPblacktext}{\newif\ifGPblacktext
    \GPblacktexttrue
  }{}\let\gplgaddtomacro\g@addto@macro
\gdef\gplbacktext{}\gdef\gplfronttext{}\makeatother
  \ifGPblacktext
\def\colorrgb#1{}\def\colorgray#1{}\else
\ifGPcolor
      \def\colorrgb#1{\color[rgb]{#1}}\def\colorgray#1{\color[gray]{#1}}\expandafter\def\csname LTw\endcsname{\color{white}}\expandafter\def\csname LTb\endcsname{\color{black}}\expandafter\def\csname LTa\endcsname{\color{black}}\expandafter\def\csname LT0\endcsname{\color[rgb]{1,0,0}}\expandafter\def\csname LT1\endcsname{\color[rgb]{0,1,0}}\expandafter\def\csname LT2\endcsname{\color[rgb]{0,0,1}}\expandafter\def\csname LT3\endcsname{\color[rgb]{1,0,1}}\expandafter\def\csname LT4\endcsname{\color[rgb]{0,1,1}}\expandafter\def\csname LT5\endcsname{\color[rgb]{1,1,0}}\expandafter\def\csname LT6\endcsname{\color[rgb]{0,0,0}}\expandafter\def\csname LT7\endcsname{\color[rgb]{1,0.3,0}}\expandafter\def\csname LT8\endcsname{\color[rgb]{0.5,0.5,0.5}}\else
\def\colorrgb#1{\color{black}}\def\colorgray#1{\color[gray]{#1}}\expandafter\def\csname LTw\endcsname{\color{white}}\expandafter\def\csname LTb\endcsname{\color{black}}\expandafter\def\csname LTa\endcsname{\color{black}}\expandafter\def\csname LT0\endcsname{\color{black}}\expandafter\def\csname LT1\endcsname{\color{black}}\expandafter\def\csname LT2\endcsname{\color{black}}\expandafter\def\csname LT3\endcsname{\color{black}}\expandafter\def\csname LT4\endcsname{\color{black}}\expandafter\def\csname LT5\endcsname{\color{black}}\expandafter\def\csname LT6\endcsname{\color{black}}\expandafter\def\csname LT7\endcsname{\color{black}}\expandafter\def\csname LT8\endcsname{\color{black}}\fi
  \fi
    \setlength{\unitlength}{0.0500bp}\ifx\gptboxheight\undefined \newlength{\gptboxheight}\newlength{\gptboxwidth}\newsavebox{\gptboxtext}\fi \setlength{\fboxrule}{0.5pt}\setlength{\fboxsep}{1pt}\definecolor{tbcol}{rgb}{1,1,1}\begin{picture}(5040.00,3440.00)\gplgaddtomacro\gplbacktext{\csname LTb\endcsname \put(735,876){\makebox(0,0)[r]{\strut{}$-200$}}\csname LTb\endcsname \put(735,1370){\makebox(0,0)[r]{\strut{}$-100$}}\csname LTb\endcsname \put(735,1864){\makebox(0,0)[r]{\strut{}$0$}}\csname LTb\endcsname \put(735,2358){\makebox(0,0)[r]{\strut{}$100$}}\csname LTb\endcsname \put(735,2852){\makebox(0,0)[r]{\strut{}$200$}}\csname LTb\endcsname \put(832,432){\makebox(0,0){\strut{}$58000$}}\csname LTb\endcsname \put(1914,432){\makebox(0,0){\strut{}$58500$}}\csname LTb\endcsname \put(2995,432){\makebox(0,0){\strut{}$59000$}}\csname LTb\endcsname \put(4077,432){\makebox(0,0){\strut{}$59500$}}}\gplgaddtomacro\gplfronttext{\csname LTb\endcsname \put(3962,2995){\makebox(0,0)[r]{\strut{}ALMA 91.5\,GHz}}\csname LTb\endcsname \put(3962,2798){\makebox(0,0)[r]{\strut{}ACT 95\,GHz (PA5)}}\csname LTb\endcsname \put(171,1926){\rotatebox{-270.00}{\makebox(0,0){\strut{}Filtered Light Curve (mJy)}}}\csname LTb\endcsname \put(2779,137){\makebox(0,0){\strut{}MJD}}}\gplbacktext
    \put(0,0){\includegraphics[width={252.00bp},height={172.00bp}]{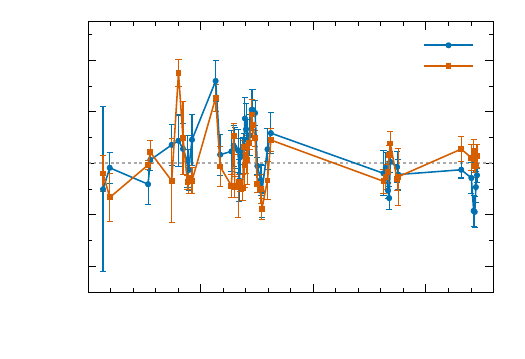}}\gplfronttext
  \end{picture}\endgroup
\\
    \begingroup
  \makeatletter
  \providecommand\color[2][]{\GenericError{(gnuplot) \space\space\space\@spaces}{Package color not loaded in conjunction with
      terminal option `colourtext'}{See the gnuplot documentation for explanation.}{Either use 'blacktext' in gnuplot or load the package
      color.sty in LaTeX.}\renewcommand\color[2][]{}}\providecommand\includegraphics[2][]{\GenericError{(gnuplot) \space\space\space\@spaces}{Package graphicx or graphics not loaded}{See the gnuplot documentation for explanation.}{The gnuplot epslatex terminal needs graphicx.sty or graphics.sty.}\renewcommand\includegraphics[2][]{}}\providecommand\rotatebox[2]{#2}\@ifundefined{ifGPcolor}{\newif\ifGPcolor
    \GPcolortrue
  }{}\@ifundefined{ifGPblacktext}{\newif\ifGPblacktext
    \GPblacktexttrue
  }{}\let\gplgaddtomacro\g@addto@macro
\gdef\gplbacktext{}\gdef\gplfronttext{}\makeatother
  \ifGPblacktext
\def\colorrgb#1{}\def\colorgray#1{}\else
\ifGPcolor
      \def\colorrgb#1{\color[rgb]{#1}}\def\colorgray#1{\color[gray]{#1}}\expandafter\def\csname LTw\endcsname{\color{white}}\expandafter\def\csname LTb\endcsname{\color{black}}\expandafter\def\csname LTa\endcsname{\color{black}}\expandafter\def\csname LT0\endcsname{\color[rgb]{1,0,0}}\expandafter\def\csname LT1\endcsname{\color[rgb]{0,1,0}}\expandafter\def\csname LT2\endcsname{\color[rgb]{0,0,1}}\expandafter\def\csname LT3\endcsname{\color[rgb]{1,0,1}}\expandafter\def\csname LT4\endcsname{\color[rgb]{0,1,1}}\expandafter\def\csname LT5\endcsname{\color[rgb]{1,1,0}}\expandafter\def\csname LT6\endcsname{\color[rgb]{0,0,0}}\expandafter\def\csname LT7\endcsname{\color[rgb]{1,0.3,0}}\expandafter\def\csname LT8\endcsname{\color[rgb]{0.5,0.5,0.5}}\else
\def\colorrgb#1{\color{black}}\def\colorgray#1{\color[gray]{#1}}\expandafter\def\csname LTw\endcsname{\color{white}}\expandafter\def\csname LTb\endcsname{\color{black}}\expandafter\def\csname LTa\endcsname{\color{black}}\expandafter\def\csname LT0\endcsname{\color{black}}\expandafter\def\csname LT1\endcsname{\color{black}}\expandafter\def\csname LT2\endcsname{\color{black}}\expandafter\def\csname LT3\endcsname{\color{black}}\expandafter\def\csname LT4\endcsname{\color{black}}\expandafter\def\csname LT5\endcsname{\color{black}}\expandafter\def\csname LT6\endcsname{\color{black}}\expandafter\def\csname LT7\endcsname{\color{black}}\expandafter\def\csname LT8\endcsname{\color{black}}\fi
  \fi
    \setlength{\unitlength}{0.0500bp}\ifx\gptboxheight\undefined \newlength{\gptboxheight}\newlength{\gptboxwidth}\newsavebox{\gptboxtext}\fi \setlength{\fboxrule}{0.5pt}\setlength{\fboxsep}{1pt}\definecolor{tbcol}{rgb}{1,1,1}\begin{picture}(5040.00,2880.00)\gplgaddtomacro\gplbacktext{\csname LTb\endcsname \put(637,957){\makebox(0,0)[r]{\strut{}$0$}}\csname LTb\endcsname \put(637,1613){\makebox(0,0)[r]{\strut{}$0.5$}}\csname LTb\endcsname \put(637,2269){\makebox(0,0)[r]{\strut{}$1$}}\csname LTb\endcsname \put(735,432){\makebox(0,0){\strut{}$10$}}\csname LTb\endcsname \put(2730,432){\makebox(0,0){\strut{}$100$}}\csname LTb\endcsname \put(4726,432){\makebox(0,0){\strut{}$1000$}}}\gplgaddtomacro\gplfronttext{\csname LTb\endcsname \put(171,1646){\rotatebox{-270.00}{\makebox(0,0){\strut{}Pearson Correlation}}}\csname LTb\endcsname \put(2730,137){\makebox(0,0){\strut{}Spline Filter Knot Spacing (days)}}}\gplbacktext
    \put(0,0){\includegraphics[width={252.00bp},height={144.00bp}]{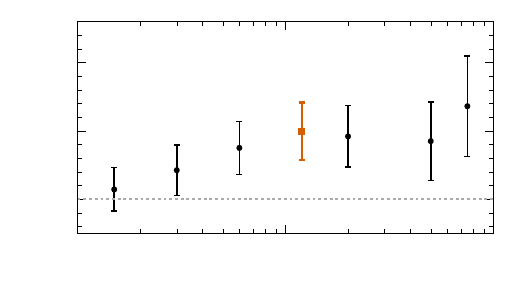}}\gplfronttext
  \end{picture}\endgroup
    \caption{Correlation of medium-term flux density variations in ALMA and ACT light curves of \pkstwo. \textit{Top:} ALMA 91.5\,GHz and ACT 95\,GHz (PA5) light curves, which have been filtered by binning and subtracting a long timescale trend. The binning is in five day chunks, and only time chunks containing data from both ALMA and ACT are retained. The long timescale trend is removed by fitting a third degree B-spline with 120-day knot spacing to each binned light curve and subtracting; this effectively acts as a high pass filter of variations on timescales $\gtrsim 120$ days, removing the large sinusoidal pattern. Clear correlations between the light curves are seen on these time scales. \textit{Bottom:} The Pearson correlation between the filtered ALMA and ACT light curves, for different spline knot spacings. Errors are estimated by calculating the correlations between the ALMA \pkstwo light curve and each of the other 204 ACT light curves in our bright AGN library, which should be uncorrelated, and taking the standard deviation. The thick, coloured point at 120 days corresponds to the data shown in the upper panel.}
   \label{plt:J2134_alma-act_corr}
\end{figure}

\subsection{Method}\label{sec:sinfit_method}

We fitted a sine wave function to the OVRO and ACT data by maximizing the following likelihood function:
\begin{align}\label{eq:mcmc_fit}
  \begin{split}
    \ln\mathcal{L} =& -\frac{1}{2}\sum_{j=1}^4\sum_{i=1}^{N_j}
    \frac{\left(S_{ij}-\bar{S}_j-A_j \sin\left[\frac{2\pi}{P}(t_{ij} - t_0)-\phi_j\right]\right)^2}{\sigma_{ij}^2+\xi_j^2} \\
    & -\frac{1}{2}\sum_{j=1}^4\sum_{i=1}^{N_j}\ln(\sigma_{ij}^2+\xi_j^2),
  \end{split}
\end{align}
where the $i$ is the time index of the light curve and $j$ stands for each of the light curves: OVRO 15\,GHz, ACT 95\,GHz, ACT 147\,GHz and ACT 225\,GHz, respectively. The light curves, $S_{ij}$, which have $N_j$ points and uncertainties $\sigma_{ij}$, are sampled at times $t_{ij}$; the reference time for the sinusoid phase is set at $t_0 =$ MJD 59\,000 (2020 May 31). The period, $P$, is held fixed during the fit (see below), and there are four free parameters per light curve: the sine-wave offset $\bar{S}_j$, amplitude $A_j$, and phase $\phi_j$, plus a term quantifying the correlated noise in the data, $\xi_j$.

We determine the period $P$ by fitting the OVRO data alone in the range where there are ACT data. Then, keeping this period fixed, we do the joint fit of Eq.~\ref{eq:mcmc_fit} described above. The reason for calculating the period only within the ${\sim}6$-year time range where there are ACT data, rather than over the whole range of the OVRO data, is that real flux fluctuations due to processes inside the jet, which are superimposed on the sinusoidal signal, act as a source of `noise' in the sine-wave fits that may not be fully captured by the correlated noise term $\xi$, and would require more SMBHB candidates to be properly understood. See O22, K25 and Appendix~\ref{app:empirical_estimate_backgnd} for a more detailed discussion of this source of uncertainty. Restricting the fit to the time range when both OVRO and ACT data exist is the best approach for measuring the phase shifts as a function of frequency (see Sec.~\ref{sec:phase_shifts}, below). A result of this restriction of the time range is that the fits with ACT and ALMA data have slightly different fixed periods ($\approx4\%$), since their time spans are not equivalent (see Fig.~\ref{plt:2131} and compare Tables~\ref{tab:2131_fits} and \ref{tab:alma_2131_fits}).

The best-fitting parameters and their uncertainties are calculated with a Markov chain Monte Carlo (MCMC) using the \textsc{emcee} code  \citep{foreman_mackey_2013}. The phase shifts are calculated from the best-fitting values as:
\begin{equation}\label{eq:phase_shift}
    \Delta\phi_j=(\phi_j-\phi_{\rm OVRO})/2\pi.
\end{equation}
The uncertainties represent the 68\% confidence range of the marginalized posterior distributions. The results of the fits are reported in Tables~\ref{tab:2131_fits} and~\ref{tab:0805_fits}; the fits to the OVRO+ALMA data, which are the same as those presented in K25 and were performed with the same methodology described above, are given in Table~\ref{tab:alma_2131_fits}.

\subsection{The light curves of \pkstwo}\label{sec:pks2131}

It is fortunate that \pkstwo\footnote{This AGN is well known in the literature as PKS~J2134$-$0153, but we use the identifier \pkstwo to be in continuity with O22 and K25.} is an ALMA calibrator that was observed during the same time period as the ACT data, as it affords us the opportunity to compare the ACT 95\,GHz light curve with ALMA at 91.5 and 103.5\,GHz, providing a useful cross-check of the calibration of the ACT data. The ACT light curves of \pkstwo at 95, 147 and 225\,GHz are shown in Fig. \ref{plt:2131}, together with the OVRO 15\,GHz and  ALMA 91.5\,GHz and 345\,GHz light curves.

The ACT 95\,GHz data show good consistency with the ALMA 91.5\,GHz and 103.5\,GHz data. We determine this by comparing the ratio of their light curves after colour-correcting the ALMA data to the ACT band centre using a spectral index of $-0.38\pm0.12$, which we obtained from the ratio of the ALMA 91.5\,GHz and 103.5\,GHz data for the 219 instances where both bands had simultaneous measurements.\footnote{In principle, the ACT band centre should also be colour-corrected, since we have adopted the band centre for synchrotron emission, i.e., an index of $-0.7$ (see Sec.~\ref{sec:actobs}). However, the difference is subdominant to other uncertainties, particularly the ${\approx}3\%$ scatter in band centre from detector array to detector array,  and we ignore it in this analysis.} The mean ratio of the 95\,GHz ACT flux to the 91.5\,ALMA flux, using the 30 pairs of data points measured on the same day in both experiments, is $1.01\pm0.04$. For ACT 95\,GHz and ALMA 103.5\,GHz, the mean ratio is $1.00\pm0.05$, using the 28 available pairs of data. Another comparison, albeit less precise, can be made using the ratio of the best-fitting sinusoid offsets, $\bar{S}_j$. After colour correcting these offsets, we find: $1.00\pm0.12$ for ACT 95\,GHz vs. ALMA 91.5\,GHz, and $1.00\pm0.13$ for ACT 95\,GHz vs. ALMA 103.5\,GHz. Note that the uncertainties here are dominated by the uncertainty in colour correction. \citet{farren/etal:2021} also found consistency between ACT and ALMA flux densities using a few dozen point sources. In their case, the comparison was between ACT fluxes averaged over a year and ALMA data from the same year, rather than between individual points in light curves as we have done here, which they note increases the measurement uncertainty. 

A further agreement between ACT and ALMA is found in the scatter in their light curves, which is similar between the two datasets on ${\sim}$month-long time scales. This `scatter' is actually due to real flux density variations in the source. We illustrate this in the top panel of Fig.~\ref{plt:J2134_alma-act_corr}, in which the long term sinusoidal variation of the ALMA and ACT light curves has been filtered out and the two resulting light curves exhibit similar residuals. The bottom panel shows the Pearson correlation between the ALMA 91.5\,GHz and ACT 95\,GHz data as a function of how aggressive this high pass filter is. On timescales $> 1$\,month the level of correlation is significant; on shorter time scales the correlation is noise dominated.

In Fig.~\ref{plt:2131} we see a monotonic progression  of the phases of the fitted sine waves with observing frequency, with  the higher frequencies leading the lower frequencies. This phenomenon was discovered by K25. \citet{ren/etal:2025} also explored phase shifts in \pkstwo using radio (OVRO), infrared (\textit{WISE}, \textit{NEOWISE}) and optical (ZTF, CRTS, Asteroid Terrestrial-impact Last Alert System (ATLAS)) light curves. They found phase shifts of $-0.19\pm0.13$\,cycles between radio and infrared and $-0.17\pm0.13$\,cycles between radio and optical.\footnote{We have converted the values they report, which are in units of days, to units of cycles using their radio period of $1760\pm33$\,days.} Since their results come from cross-correlation the uncertainties are much larger than K25 but are consistent within the errorbars. We analyse the phase shifts later, in Sec.~\ref{sec:phase_shifts}.

\subsection{The ACT light curve of \pkszero}\label{sec:pks0805}

\begin{figure*}[tb]
   \centering
   \begingroup
  \makeatletter
  \providecommand\color[2][]{\GenericError{(gnuplot) \space\space\space\@spaces}{Package color not loaded in conjunction with
      terminal option `colourtext'}{See the gnuplot documentation for explanation.}{Either use 'blacktext' in gnuplot or load the package
      color.sty in LaTeX.}\renewcommand\color[2][]{}}\providecommand\includegraphics[2][]{\GenericError{(gnuplot) \space\space\space\@spaces}{Package graphicx or graphics not loaded}{See the gnuplot documentation for explanation.}{The gnuplot epslatex terminal needs graphicx.sty or graphics.sty.}\renewcommand\includegraphics[2][]{}}\providecommand\rotatebox[2]{#2}\@ifundefined{ifGPcolor}{\newif\ifGPcolor
    \GPcolortrue
  }{}\@ifundefined{ifGPblacktext}{\newif\ifGPblacktext
    \GPblacktexttrue
  }{}\let\gplgaddtomacro\g@addto@macro
\gdef\gplbacktext{}\gdef\gplfronttext{}\makeatother
  \ifGPblacktext
\def\colorrgb#1{}\def\colorgray#1{}\else
\ifGPcolor
      \def\colorrgb#1{\color[rgb]{#1}}\def\colorgray#1{\color[gray]{#1}}\expandafter\def\csname LTw\endcsname{\color{white}}\expandafter\def\csname LTb\endcsname{\color{black}}\expandafter\def\csname LTa\endcsname{\color{black}}\expandafter\def\csname LT0\endcsname{\color[rgb]{1,0,0}}\expandafter\def\csname LT1\endcsname{\color[rgb]{0,1,0}}\expandafter\def\csname LT2\endcsname{\color[rgb]{0,0,1}}\expandafter\def\csname LT3\endcsname{\color[rgb]{1,0,1}}\expandafter\def\csname LT4\endcsname{\color[rgb]{0,1,1}}\expandafter\def\csname LT5\endcsname{\color[rgb]{1,1,0}}\expandafter\def\csname LT6\endcsname{\color[rgb]{0,0,0}}\expandafter\def\csname LT7\endcsname{\color[rgb]{1,0.3,0}}\expandafter\def\csname LT8\endcsname{\color[rgb]{0.5,0.5,0.5}}\else
\def\colorrgb#1{\color{black}}\def\colorgray#1{\color[gray]{#1}}\expandafter\def\csname LTw\endcsname{\color{white}}\expandafter\def\csname LTb\endcsname{\color{black}}\expandafter\def\csname LTa\endcsname{\color{black}}\expandafter\def\csname LT0\endcsname{\color{black}}\expandafter\def\csname LT1\endcsname{\color{black}}\expandafter\def\csname LT2\endcsname{\color{black}}\expandafter\def\csname LT3\endcsname{\color{black}}\expandafter\def\csname LT4\endcsname{\color{black}}\expandafter\def\csname LT5\endcsname{\color{black}}\expandafter\def\csname LT6\endcsname{\color{black}}\expandafter\def\csname LT7\endcsname{\color{black}}\expandafter\def\csname LT8\endcsname{\color{black}}\fi
  \fi
    \setlength{\unitlength}{0.0500bp}\ifx\gptboxheight\undefined \newlength{\gptboxheight}\newlength{\gptboxwidth}\newsavebox{\gptboxtext}\fi \setlength{\fboxrule}{0.5pt}\setlength{\fboxsep}{1pt}\definecolor{tbcol}{rgb}{1,1,1}\begin{picture}(10080.00,5320.00)\gplgaddtomacro\gplbacktext{\csname LTb\endcsname \put(685,530){\makebox(0,0)[r]{\strut{}$0$}}\csname LTb\endcsname \put(685,999){\makebox(0,0)[r]{\strut{}$0.1$}}\csname LTb\endcsname \put(685,1467){\makebox(0,0)[r]{\strut{}$0.2$}}\csname LTb\endcsname \put(685,1936){\makebox(0,0)[r]{\strut{}$0.3$}}\csname LTb\endcsname \put(685,2405){\makebox(0,0)[r]{\strut{}$0.4$}}\csname LTb\endcsname \put(685,2873){\makebox(0,0)[r]{\strut{}$0.5$}}\csname LTb\endcsname \put(685,3342){\makebox(0,0)[r]{\strut{}$0.6$}}\csname LTb\endcsname \put(685,3810){\makebox(0,0)[r]{\strut{}$0.7$}}\csname LTb\endcsname \put(685,4279){\makebox(0,0)[r]{\strut{}$0.8$}}\csname LTb\endcsname \put(783,393){\makebox(0,0){\strut{}$57600$}}\csname LTb\endcsname \put(1842,393){\makebox(0,0){\strut{}$57800$}}\csname LTb\endcsname \put(2901,393){\makebox(0,0){\strut{}$58000$}}\csname LTb\endcsname \put(3960,393){\makebox(0,0){\strut{}$58200$}}\csname LTb\endcsname \put(5019,393){\makebox(0,0){\strut{}$58400$}}\csname LTb\endcsname \put(6078,393){\makebox(0,0){\strut{}$58600$}}\csname LTb\endcsname \put(7137,393){\makebox(0,0){\strut{}$58800$}}\csname LTb\endcsname \put(8196,393){\makebox(0,0){\strut{}$59000$}}\csname LTb\endcsname \put(9255,393){\makebox(0,0){\strut{}$59200$}}\csname LTb\endcsname \put(1598,4651){\makebox(0,0){\strut{}2017}}\csname LTb\endcsname \put(3531,4651){\makebox(0,0){\strut{}2018}}\csname LTb\endcsname \put(5464,4651){\makebox(0,0){\strut{}2019}}\csname LTb\endcsname \put(7397,4651){\makebox(0,0){\strut{}2020}}\csname LTb\endcsname \put(9335,4651){\makebox(0,0){\strut{}2021}}}\gplgaddtomacro\gplfronttext{\csname LTb\endcsname \put(9286,4336){\makebox(0,0)[r]{\strut{}OVRO 15\,GHz}}\csname LTb\endcsname \put(9286,4140){\makebox(0,0)[r]{\strut{}ACT 95\,GHz}}\csname LTb\endcsname \put(9286,3943){\makebox(0,0)[r]{\strut{}ACT 147\,GHz}}\csname LTb\endcsname \put(9286,3746){\makebox(0,0)[r]{\strut{}ACT 225\,GHz}}\csname LTb\endcsname \put(268,2522){\rotatebox{-270.00}{\makebox(0,0){\strut{}Flux Density (Jy)}}}\csname LTb\endcsname \put(5416,137){\makebox(0,0){\strut{}Modified Julian Day}}\csname LTb\endcsname \put(5416,5005){\makebox(0,0){\strut{}\textbf{\pkszero}}}}\gplgaddtomacro\gplbacktext{}\gplgaddtomacro\gplfronttext{\csname LTb\endcsname \put(1209,2597){\makebox(0,0)[r]{\strut{}\footnotesize 0}}\csname LTb\endcsname \put(1209,3015){\makebox(0,0)[r]{\strut{}\footnotesize 0.2}}\csname LTb\endcsname \put(1209,3433){\makebox(0,0)[r]{\strut{}\footnotesize 0.4}}\csname LTb\endcsname \put(1209,3851){\makebox(0,0)[r]{\strut{}\footnotesize 0.6}}\csname LTb\endcsname \put(1667,4263){\makebox(0,0){\strut{}2010}}\csname LTb\endcsname \put(2326,4263){\makebox(0,0){\strut{}2014}}\csname LTb\endcsname \put(2984,4263){\makebox(0,0){\strut{}2018}}\csname LTb\endcsname \put(3643,4263){\makebox(0,0){\strut{}2022}}}\gplbacktext
    \put(0,0){\includegraphics[width={504.00bp},height={266.00bp}]{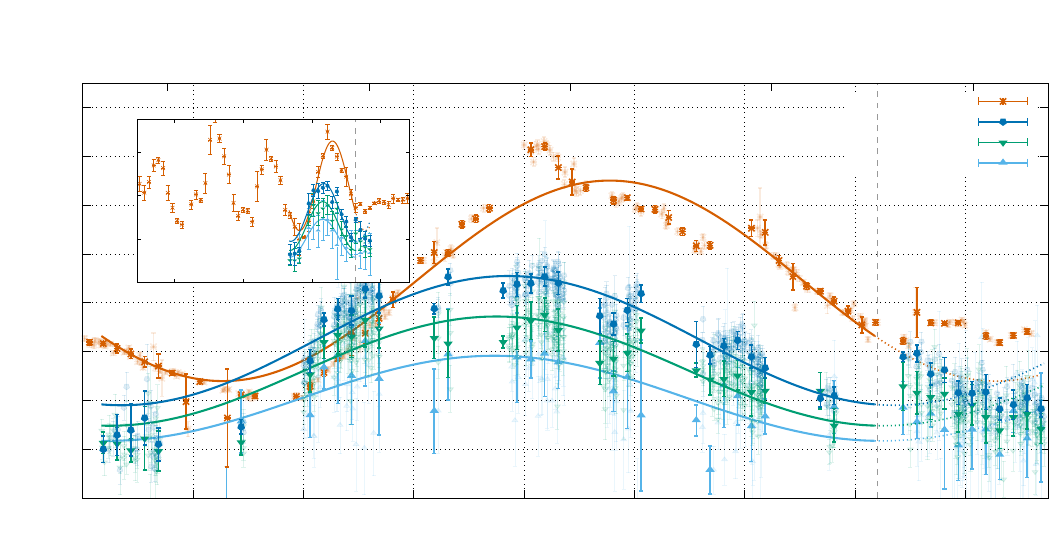}}\gplfronttext
  \end{picture}\endgroup
 \caption{\pkszero light curves at 15\,GHz, 95\,GHz,  147\,GHz, and 225\,GHz. The main panel shows the time range covered by the ACT observations while the inset shows the full range observed by OVRO. Heavy points are binned into 25-day intervals (100 days in the inset) to guide the eye, with individual measurements shown as lighter points; the sinusoidal, least-squares fits were done on the unbinned data. A monotonic phase shift of the sine waves to earlier times with increasing frequency is visible by eye, and is reported relative to the OVRO phase as $\Delta\phi_0$ in Table~\ref{tab:0805_fits}. The dashed vertical line at MJD 59041 (2020 July 11) indicates the approximate date at which D25 found that the sinusoidal variation in the 15\,GHz ceases. The ACT mm data also apparently flatten around the same time.}
         \label{plt:0805}
\end{figure*}

\begin{table*}[h]
\centering
\caption{Sine-wave fit results for \pkszero.}
\begin{tabular}{lrrrr}
\hline \hline
Parameter & OVRO 15 GHz & ACT 95 GHz & ACT 147 GHz & ACT 225 GHz \\
\hline
$P$ (days)              & \multicolumn{4}{c}{$1402\pm23$ (fixed)} \\
$A$   (Jy)              & $0.2054 \pm 0.0057$ & $0.1324 \pm 0.0033$  & $0.1119 \pm 0.0030$  & $0.0872 \pm 0.0092$ \\
$\phi_0$ (rad)          & $2.721 \pm 0.028$   & $1.896 \pm 0.022$    & $1.800 \pm 0.024$    & $1.784 \pm 0.089$ \\
$\bar{S}$ (Jy)          & $0.4446 \pm 0.0041$ & $0.3221 \pm 0.0022$  & $0.2596 \pm 0.0020$  & $0.2039 \pm 0.0058$ \\
$\xi$ (Jy)              & $0.0482 \pm 0.0030$ & $0.0410 \pm 0.0018$  & $0.0408 \pm 0.0018$  & $0.0412 \pm 0.0076$ \\
$\Delta\phi_0$ (cycles) & 0                   & $-0.1312 \pm 0.0056$ & $-0.1465 \pm 0.0058$ & $-0.1490 \pm 0.0148$ \\
$\Delta\phi_0$ (days)   & 0                   & $-184.0 \pm 7.9$     & $-205.5 \pm 8.1$     & $-208.9 \pm 20.8$ \\
\hline
\end{tabular}
\tablefoot{The period, $P$, was determined from an initial fit including only the OVRO data in the range $57634 < \mathrm{MJD} < 59343$. The joint fit including OVRO and the three ACT light curves kept the period fixed at this best fitting value. The fit uncertainties do not account for systematics due to superimposed, shorter term variations (see Appendix~\ref{app:empirical_estimate_backgnd}). The reference time for the phase, in MJD, is  $t_0 = 59\,000$. The phase shift, $\Delta\phi_0$, is defined in Eq.~\ref{eq:phase_shift}.}
\label{tab:0805_fits}
\end{table*}

We now turn to the case of \pkszero, for which there are no ALMA data. This is the second OVRO blazar discovered to exhibit significant sinusoidal variations, thereby making it a strong SMBHB candidate (D25). Best-fitting sinusoids to the ACT and OVRO data are reported in Table~\ref{tab:0805_fits} and are shown along with the light curves in Fig.~\ref{plt:0805}.  The sinusoidal variation discovered at 15\,GHz in the OVRO light curve of \pkszero, which lasted from 2008 to 2020 (D25), is clearly also seen in the ACT light curves, and shows that the broadband sinusoidal variation seen in \pkstwo is not simply an isolated case. It suggests that broadband sinusoidal emission is a common phenomenon in SMBHB blazar candidates. The KO model explains this phenomenology which has now been confirmed in our two, high significance SMBHB candidates. This could be important for the detection of gravitational waves from SMBHBs.

D25 found that the sinusoidal variations in the 15\,GHz OVRO light curve ceased at around MJD~59041, shown in Fig.~\ref{plt:0805} with a dashed, vertical line. Although the ACT light curves do not extend long after this date, they too appear consistent with the sinusoid shutting off. A similar phenomenon was also observed by O22 and K25 in \pkstwo in more than 45 years of data: its light curve was sinusoidal for about the first seven years, then non-sinusoidal for 19 years, after which the sinusoid recommenced with a smaller amplitude but otherwise consistent, in both period and phase, with the original periodicity. Completely independent fits to the Haystack data (1975--1983) and to the OVRO data (2008--2020) gave periods of $1729.1\pm32.4$ days and $1760\pm5.3$ days, respectively (O22). This is the same period within the uncertainties, which is better than 2\%. Moreover, the new sinusoid was in phase with the original to within ${\sim}10\%$ of a cycle (O22). This suggests that the underlying mechanism is a good clock---i.e., the SMBHB orbit according to the KO model---that continued running when the  sinusoidal variation seen in the Haystack and OVRO data was absent. The temporary cessation of the sinusoid is not explained by KO model \textit{per se}, but O22 and K25 point out that its disappearance, as well as the change in its amplitude when it reappears, could plausibly be produced by changes in the fuelling of the jet. We discuss this intermittency in more detail in Sec.~\ref{ssec:bh_sims}.

As in the case of \pkstwo, a monotonic phase shift with frequency is observed in the light curves, which we analyse in Sec.~\ref{sec:phase_shifts}, below.

\section{Multi-frequency properties of the SMBHB candidates}\label{sec:multifreq_properties}

\subsection{Frequency-dependent phase shifts}\label{sec:phase_shifts}

\begin{figure}[tb]
   \centering
\begingroup
  \makeatletter
  \providecommand\color[2][]{\GenericError{(gnuplot) \space\space\space\@spaces}{Package color not loaded in conjunction with
      terminal option `colourtext'}{See the gnuplot documentation for explanation.}{Either use 'blacktext' in gnuplot or load the package
      color.sty in LaTeX.}\renewcommand\color[2][]{}}\providecommand\includegraphics[2][]{\GenericError{(gnuplot) \space\space\space\@spaces}{Package graphicx or graphics not loaded}{See the gnuplot documentation for explanation.}{The gnuplot epslatex terminal needs graphicx.sty or graphics.sty.}\renewcommand\includegraphics[2][]{}}\providecommand\rotatebox[2]{#2}\@ifundefined{ifGPcolor}{\newif\ifGPcolor
    \GPcolortrue
  }{}\@ifundefined{ifGPblacktext}{\newif\ifGPblacktext
    \GPblacktexttrue
  }{}\let\gplgaddtomacro\g@addto@macro
\gdef\gplbacktext{}\gdef\gplfronttext{}\makeatother
  \ifGPblacktext
\def\colorrgb#1{}\def\colorgray#1{}\else
\ifGPcolor
      \def\colorrgb#1{\color[rgb]{#1}}\def\colorgray#1{\color[gray]{#1}}\expandafter\def\csname LTw\endcsname{\color{white}}\expandafter\def\csname LTb\endcsname{\color{black}}\expandafter\def\csname LTa\endcsname{\color{black}}\expandafter\def\csname LT0\endcsname{\color[rgb]{1,0,0}}\expandafter\def\csname LT1\endcsname{\color[rgb]{0,1,0}}\expandafter\def\csname LT2\endcsname{\color[rgb]{0,0,1}}\expandafter\def\csname LT3\endcsname{\color[rgb]{1,0,1}}\expandafter\def\csname LT4\endcsname{\color[rgb]{0,1,1}}\expandafter\def\csname LT5\endcsname{\color[rgb]{1,1,0}}\expandafter\def\csname LT6\endcsname{\color[rgb]{0,0,0}}\expandafter\def\csname LT7\endcsname{\color[rgb]{1,0.3,0}}\expandafter\def\csname LT8\endcsname{\color[rgb]{0.5,0.5,0.5}}\else
\def\colorrgb#1{\color{black}}\def\colorgray#1{\color[gray]{#1}}\expandafter\def\csname LTw\endcsname{\color{white}}\expandafter\def\csname LTb\endcsname{\color{black}}\expandafter\def\csname LTa\endcsname{\color{black}}\expandafter\def\csname LT0\endcsname{\color{black}}\expandafter\def\csname LT1\endcsname{\color{black}}\expandafter\def\csname LT2\endcsname{\color{black}}\expandafter\def\csname LT3\endcsname{\color{black}}\expandafter\def\csname LT4\endcsname{\color{black}}\expandafter\def\csname LT5\endcsname{\color{black}}\expandafter\def\csname LT6\endcsname{\color{black}}\expandafter\def\csname LT7\endcsname{\color{black}}\expandafter\def\csname LT8\endcsname{\color{black}}\fi
  \fi
    \setlength{\unitlength}{0.0500bp}\ifx\gptboxheight\undefined \newlength{\gptboxheight}\newlength{\gptboxwidth}\newsavebox{\gptboxtext}\fi \setlength{\fboxrule}{0.5pt}\setlength{\fboxsep}{1pt}\definecolor{tbcol}{rgb}{1,1,1}\begin{picture}(4680.00,5040.00)\gplgaddtomacro\gplbacktext{\csname LTb\endcsname \put(734,2224){\makebox(0,0)[r]{\strut{} $-0.3$}}\csname LTb\endcsname \put(734,2744){\makebox(0,0)[r]{\strut{} $-0.2$}}\csname LTb\endcsname \put(734,3264){\makebox(0,0)[r]{\strut{} $-0.1$}}\csname LTb\endcsname \put(734,3783){\makebox(0,0)[r]{\strut{} $0$}}\csname LTb\endcsname \put(734,4303){\makebox(0,0)[r]{\strut{} $0.1$}}\csname LTb\endcsname \put(734,4823){\makebox(0,0)[r]{\strut{} $0.2$}}\csname LTb\endcsname \put(832,1560){\makebox(0,0){\strut{}}}\csname LTb\endcsname \put(1331,1560){\makebox(0,0){\strut{}}}\csname LTb\endcsname \put(1830,1560){\makebox(0,0){\strut{}}}\csname LTb\endcsname \put(2329,1560){\makebox(0,0){\strut{}}}\csname LTb\endcsname \put(2829,1560){\makebox(0,0){\strut{}}}\csname LTb\endcsname \put(3328,1560){\makebox(0,0){\strut{}}}\csname LTb\endcsname \put(3827,1560){\makebox(0,0){\strut{}}}\csname LTb\endcsname \put(3925,2067){\makebox(0,0)[l]{\strut{}$-600$}}\csname LTb\endcsname \put(3925,2639){\makebox(0,0)[l]{\strut{}$-400$}}\csname LTb\endcsname \put(3925,3211){\makebox(0,0)[l]{\strut{}$-200$}}\csname LTb\endcsname \put(3925,3783){\makebox(0,0)[l]{\strut{}$0$}}\csname LTb\endcsname \put(3925,4356){\makebox(0,0)[l]{\strut{}$200$}}}\gplgaddtomacro\gplfronttext{\csname LTb\endcsname \put(3259,4253){\makebox(0,0)[r]{\strut{}Power Law}}\csname LTb\endcsname \put(3259,4449){\makebox(0,0)[r]{\strut{}K24}}\csname LTb\endcsname \put(3259,4646){\makebox(0,0)[r]{\strut{}ACT}}\csname LTb\endcsname \put(72,3290){\rotatebox{-270.00}{\makebox(0,0){\strut{}Phase Shift (cycles)}}}\csname LTb\endcsname \put(4562,3290){\rotatebox{-270.00}{\makebox(0,0){\strut{}Phase Shift (days)}}}\csname LTb\endcsname \put(2329,1501){\makebox(0,0){\strut{}}}}\gplgaddtomacro\gplbacktext{\csname LTb\endcsname \put(734,629){\makebox(0,0)[r]{\strut{}$-0.05$}}\csname LTb\endcsname \put(734,992){\makebox(0,0)[r]{\strut{}$0$}}\csname LTb\endcsname \put(734,1356){\makebox(0,0)[r]{\strut{}$0.05$}}\csname LTb\endcsname \put(734,1720){\makebox(0,0)[r]{\strut{}$0.1$}}\csname LTb\endcsname \put(832,432){\makebox(0,0){\strut{}$10^{0}$}}\csname LTb\endcsname \put(1331,432){\makebox(0,0){\strut{}$10^{1}$}}\csname LTb\endcsname \put(1830,432){\makebox(0,0){\strut{}$10^{2}$}}\csname LTb\endcsname \put(2329,432){\makebox(0,0){\strut{}$10^{3}$}}\csname LTb\endcsname \put(2829,432){\makebox(0,0){\strut{}$10^{4}$}}\csname LTb\endcsname \put(3328,432){\makebox(0,0){\strut{}$10^{5}$}}\csname LTb\endcsname \put(3827,432){\makebox(0,0){\strut{}$10^{6}$}}\csname LTb\endcsname \put(3925,992){\makebox(0,0)[l]{\strut{}$0$ }}\csname LTb\endcsname \put(3925,1393){\makebox(0,0)[l]{\strut{}$100$ }}}\gplgaddtomacro\gplfronttext{\csname LTb\endcsname \put(72,1193){\rotatebox{-270.00}{\makebox(0,0){\strut{}Resid.}}}\csname LTb\endcsname \put(4562,1193){\rotatebox{-270.00}{\makebox(0,0){\strut{}Resid.}}}\csname LTb\endcsname \put(2329,137){\makebox(0,0){\strut{}Frequency (GHz)}}}\gplbacktext
    \put(0,0){\includegraphics[width={234.00bp},height={252.00bp}]{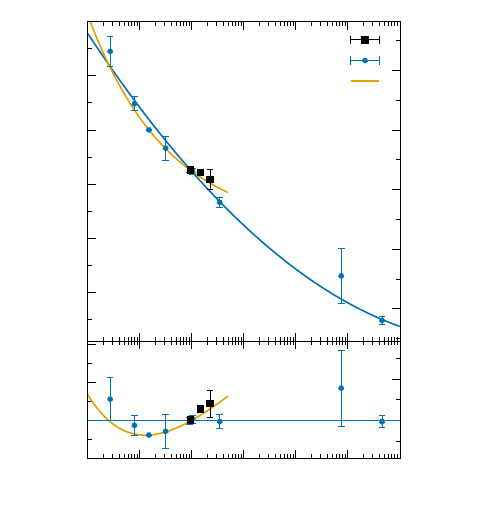}}\gplfronttext
  \end{picture}\endgroup
       \caption{\pkstwo phase shifts measured by ACT at 95\,GHz, 147\,GHz, and 225\,GHz (black points), compared to the phase shift results of K25 (blue points); see Table~\ref{tab:2131_phases}. Note that the ACT 95\,GHz and ALMA\,91.5\,GHz points overlap. The Haystack or OVRO 15 GHz light curves provided the phase reference in all cases. Uncertainties are determined by the MCMC sine-wave fits (Eq.~\ref{eq:mcmc_fit}). The curved, blue line shows the quadratic fit determined by K25 (Eq.~\ref{eq:2131_quadratic_fit}). The orange line is the best fit power law to the Haystack, ALMA and ACT data (Eq.~\ref{eq:powlaw}, Table~\ref{tab:phase_comparison}).}
         \label{plt:2131_phases}
\end{figure}

\begin{table}[tb]
  \centering
  \caption{\pkstwo Phase Shifts with Frequency}
  \begin{tabular}{cccc}
    \hline\hline
    Instrument & Frequency& Phase& Uncert.\\
    & Band & (Fraction & (Fraction \\
    & & of cycle)& of cycle)\\
    \hline
    Haystack      & 2.7\,GHz  & 0.145&0.028\\
    Haystack      & 7.9\,GHz  & 0.048&0.013\\
    Haystack/OVRO & 15\,GHz\tablefootmark{a}   & 0&0\\
    Haystack      & 31.4\,GHz & $-0.034$&0.022\\
    ALMA          & 91.5\,GHz & $-0.073$&0.005\\
    ACT           & 95\,GHz   & $-0.074$&0.005\\
    ALMA          & 103.5\,GHz  & $-0.077$&0.005\\
    ACT           & 147\,GHz  & $-0.079$&0.005\\
    ACT           & 225\,GHz  & $-0.092$&0.018\\
    ALMA          & 345\,GHz  & $-0.134$&0.009\\
    \textit{WISE} & infrared\tablefootmark{b}  & $-0.27$&0.05\\
    ZTF           & optical\tablefootmark{c}   & $-0.352$&0.008\\
    \hline
  \end{tabular}
  \tablefoot{The OVRO light curve was the phase reference, except for the Haystack data, whose light curves do no overlap with the others, and which use the Haystack 15.5\,GHz channel as the reference. Apart from the ACT results, data in this table are taken directly from table 2 of K25.\\
  \tablefoottext{a}{The Haystack channel is 15.5\,GHz; OVRO is 15\,GHz.}
  \tablefoottext{b}{Both the \textit{WISE} 1 (2.8--3.8\,$\si{\micro\meter}$) and \textit{WISE} 2 (4.1--5.2\,$\si{\micro\meter})$ bands were analysed, and gave the same phase shift.}
  \tablefoottext{c}{ZTF phase shift is from the combined $r$-, $g$- and $i$-bands.}}
  \label{tab:2131_phases}
\end{table}

\begin{table}[tb]
  \centering
  \caption{\pkszero Phase Shifts with Frequency Relative to the 15\,GHz  Light Curve }
  \begin{tabular}{ccc}
    \hline\hline
    Frequency&Phase&Uncertainty\\
    Band&(Fraction&(Fraction\\
    &of cycle)&of cycle)\\
    \hline
    15\,GHz&0&0\\
    95\,GHz&$-0.131$&0.006\\
    147\,GHz&$-0.147$&0.006\\
    225\,GHz&$-0.149$&0.015\\
    \hline
  \end{tabular}
  \tablefoot{The  15~GHz light curve was used as the phase reference for the ACT data.}
  \label{tab:0805_phases}
\end{table}

As noted in the previous section, in both of our sources the higher frequency ACT light curves are shifted towards earlier times relative to the OVRO 15\,GHz light curves, and the shift is monotonic with frequency. This behaviour is explained by the KO model as being due to frequencies originating at different positions along the jet because of  optical depth effects, and is further discussed below in the context of our new MKO model (Sec.~\ref{ssec:mko}).

Table~\ref{tab:2131_phases} shows the phase shifts for \pkstwo. In addition to the phase shifts of the ACT light curves relative to OVRO, it includes the phase shifts of light curves from the Haystack observatory (radio), ALMA (mm), the Wide-field Infrared Survey Explorer (\textit{WISE}; infrared) and ZTF (optical), all relative to OVRO, as listed in table 2 of K25. Note that the Haystack data came from an earlier time period (1975--1983) and did not overlap with the other datasets; thus, the phase shift for the Haystack light curves are relative to the 15.5\,GHz Haystack channel. The coherent behaviour from 2.7 GHz to optical wavelengths is strikingly illustrated by Fig.~\ref{plt:2131_phases}, which shows the phase shift measurements on top of the best-fit quadratic phase--frequency relation from K25,
\begin{equation}\label{eq:2131_quadratic_fit}
    \Delta \phi = 0.178 - 0.146\log_{10}(\nu) + 0.0093[\log_{10}(\nu)]^2.
\end{equation}    
We discuss these phase shifts and Eq.~\ref{eq:2131_quadratic_fit} shortly, below.

Table~\ref{tab:0805_phases} lists the phase shifts for \pkszero. Note that, as found in \pkstwo, the ACT sinusoids are shifted towards earlier times relative to the OVRO 15\,GHz light curve, and the shift is consistent with being monotonic with frequency.
This is therefore likely to be a common phenomenology in SMBHB blazar candidates and a signature of a fundamental property of these jets.

\begin{table*}
  \centering
  \caption{Power law fits to the frequency--phase relation (Eq.~\ref{eq:powlaw}) at select ranges of radio and mm frequencies}
  \renewcommand{\arraystretch}{1.25}
  \hspace{-2em}\begin{tabular}{|c|cc|cc|cc|}
      \hline
      AGN & Data Sets & Span & $b$ & $a$  & $\chi^2$/dof & PTE \\
      & & (GHz) & & (cycles\,GHz$^{-1}$) & & \\
      \hline
      \pkstwo  & Haystack, OVRO, ALMA, ACT & 3--345 & $-0.3^{+0.1}_{-0.1}$ & $0.396^{+0.09}_{-0.001}$ & $13.3/7$ & 0.07\\
      \pkstwo  & Haystack, OVRO, ALMA, ACT & 3--225 & $-0.46^{+0.1}_{-0.09}$ & $0.44^{+0.06}_{-0.04}$ & $1.01/6$ & 0.99\\
      \pkstwo  & OVRO, ACT            & 15--225 & $-0.8^{+0.1}_{-0.6}$ & $0.8^{+2}_{-0.2}$ & $0.21/1$ & 0.65\\
      \pkszero & OVRO, ACT            & 15--225 & $-0.7^{+0.2}_{-0.4}$ & $1.2^{+2}_{-0.3}$ & $0.22/1$ & 0.64\\
      \hline
  \end{tabular}
  \tablefoot{Uncertainties are 68\% credible intervals.}
  \label{tab:phase_comparison}
\end{table*}

\begin{figure}[tb]
  \centering
  \footnotesize
  \begingroup
  \makeatletter
  \providecommand\color[2][]{\GenericError{(gnuplot) \space\space\space\@spaces}{Package color not loaded in conjunction with
      terminal option `colourtext'}{See the gnuplot documentation for explanation.}{Either use 'blacktext' in gnuplot or load the package
      color.sty in LaTeX.}\renewcommand\color[2][]{}}\providecommand\includegraphics[2][]{\GenericError{(gnuplot) \space\space\space\@spaces}{Package graphicx or graphics not loaded}{See the gnuplot documentation for explanation.}{The gnuplot epslatex terminal needs graphicx.sty or graphics.sty.}\renewcommand\includegraphics[2][]{}}\providecommand\rotatebox[2]{#2}\@ifundefined{ifGPcolor}{\newif\ifGPcolor
    \GPcolortrue
  }{}\@ifundefined{ifGPblacktext}{\newif\ifGPblacktext
    \GPblacktexttrue
  }{}\let\gplgaddtomacro\g@addto@macro
\gdef\gplbacktext{}\gdef\gplfronttext{}\makeatother
  \ifGPblacktext
\def\colorrgb#1{}\def\colorgray#1{}\else
\ifGPcolor
      \def\colorrgb#1{\color[rgb]{#1}}\def\colorgray#1{\color[gray]{#1}}\expandafter\def\csname LTw\endcsname{\color{white}}\expandafter\def\csname LTb\endcsname{\color{black}}\expandafter\def\csname LTa\endcsname{\color{black}}\expandafter\def\csname LT0\endcsname{\color[rgb]{1,0,0}}\expandafter\def\csname LT1\endcsname{\color[rgb]{0,1,0}}\expandafter\def\csname LT2\endcsname{\color[rgb]{0,0,1}}\expandafter\def\csname LT3\endcsname{\color[rgb]{1,0,1}}\expandafter\def\csname LT4\endcsname{\color[rgb]{0,1,1}}\expandafter\def\csname LT5\endcsname{\color[rgb]{1,1,0}}\expandafter\def\csname LT6\endcsname{\color[rgb]{0,0,0}}\expandafter\def\csname LT7\endcsname{\color[rgb]{1,0.3,0}}\expandafter\def\csname LT8\endcsname{\color[rgb]{0.5,0.5,0.5}}\else
\def\colorrgb#1{\color{black}}\def\colorgray#1{\color[gray]{#1}}\expandafter\def\csname LTw\endcsname{\color{white}}\expandafter\def\csname LTb\endcsname{\color{black}}\expandafter\def\csname LTa\endcsname{\color{black}}\expandafter\def\csname LT0\endcsname{\color{black}}\expandafter\def\csname LT1\endcsname{\color{black}}\expandafter\def\csname LT2\endcsname{\color{black}}\expandafter\def\csname LT3\endcsname{\color{black}}\expandafter\def\csname LT4\endcsname{\color{black}}\expandafter\def\csname LT5\endcsname{\color{black}}\expandafter\def\csname LT6\endcsname{\color{black}}\expandafter\def\csname LT7\endcsname{\color{black}}\expandafter\def\csname LT8\endcsname{\color{black}}\fi
  \fi
    \setlength{\unitlength}{0.0500bp}\ifx\gptboxheight\undefined \newlength{\gptboxheight}\newlength{\gptboxwidth}\newsavebox{\gptboxtext}\fi \setlength{\fboxrule}{0.5pt}\setlength{\fboxsep}{1pt}\definecolor{tbcol}{rgb}{1,1,1}\begin{picture}(5040.00,3600.00)\gplgaddtomacro\gplbacktext{\csname LTb\endcsname \put(637,629){\makebox(0,0)[r]{\strut{}$0$}}\csname LTb\endcsname \put(637,973){\makebox(0,0)[r]{\strut{}$0.5$}}\csname LTb\endcsname \put(637,1317){\makebox(0,0)[r]{\strut{}$1$}}\csname LTb\endcsname \put(637,1661){\makebox(0,0)[r]{\strut{}$1.5$}}\csname LTb\endcsname \put(637,2006){\makebox(0,0)[r]{\strut{}$2$}}\csname LTb\endcsname \put(637,2350){\makebox(0,0)[r]{\strut{}$2.5$}}\csname LTb\endcsname \put(637,2694){\makebox(0,0)[r]{\strut{}$3$}}\csname LTb\endcsname \put(637,3039){\makebox(0,0)[r]{\strut{}$3.5$}}\csname LTb\endcsname \put(637,3383){\makebox(0,0)[r]{\strut{}$4$}}\csname LTb\endcsname \put(1162,432){\makebox(0,0){\strut{}$-1.25$}}\csname LTb\endcsname \put(1875,432){\makebox(0,0){\strut{}$-1$}}\csname LTb\endcsname \put(2588,432){\makebox(0,0){\strut{}$-0.75$}}\csname LTb\endcsname \put(3300,432){\makebox(0,0){\strut{}$-0.5$}}\csname LTb\endcsname \put(4013,432){\makebox(0,0){\strut{}$-0.25$}}\csname LTb\endcsname \put(4726,432){\makebox(0,0){\strut{}$0$}}}\gplgaddtomacro\gplfronttext{\csname LTb\endcsname \put(2504,3190){\makebox(0,0)[l]{\strut{}PKS~2131, 3--345\,GHz}}\csname LTb\endcsname \put(2504,2993){\makebox(0,0)[l]{\strut{}PKS~2131, 3--225\,GHz}}\csname LTb\endcsname \put(2504,2796){\makebox(0,0)[l]{\strut{}PKS~2131, 15--225\,GHz}}\csname LTb\endcsname \put(2504,2600){\makebox(0,0)[l]{\strut{}PKS~J0805, 15--225\,GHz}}\csname LTb\endcsname \put(220,2006){\rotatebox{-270.00}{\makebox(0,0){\strut{}Amplitude $a$ (cycles\,GHz$^{-1}$)}}}\csname LTb\endcsname \put(2730,137){\makebox(0,0){\strut{}Index $b$}}}\gplbacktext
    \put(0,0){\includegraphics[width={252.00bp},height={180.00bp}]{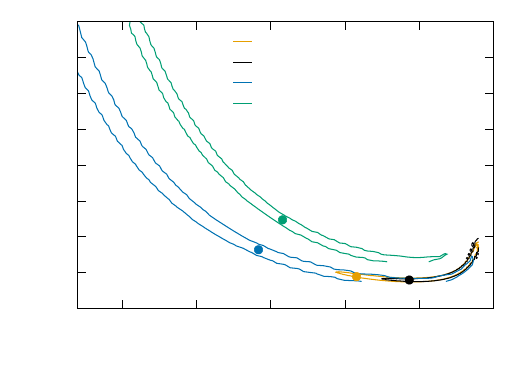}}\gplfronttext
  \end{picture}\endgroup
   \caption{Posterior probabilities of the power law MC fit to the sinusoid phase shifts (Eq.~\ref{eq:powlaw}) for different data combinations. The contours show the 95\% credible regions and the dots are the maximum likelihood values (c.f., Table~\ref{tab:phase_comparison}). Flat priors on $a$ and $b$ in the ranges of $(0.05, 4)$ and ($-2, -0.05$) were used.}
  \label{plt:phase_mcmc}
\end{figure}

The exact form of the frequency phase--shift relation depends on the details of  where different frequencies originate in the blazar jet. The simple quadratic fit by K25 to the \pkstwo phase shifts (Eq.~\ref{eq:2131_quadratic_fit}) is not motivated by any particular jet model but is purely phenomenological.  While the ACT 95\,GHz measurement agrees with this curve, and is consistent with the ALMA measurement, the 147\,GHz and 225\,GHz measurements are $3.1\sigma$ and $1.2\sigma$ higher than the curve, respectively. A power law is potentially more physically realistic:
\begin{equation}\label{eq:powlaw}
  \Delta\phi = a (\nu^b - \nu_0^b),
\end{equation}
where $a$ and $b$ are model parameters and $\nu_0$ is the reference frequency for the phase shifts, in our case 15\,GHz. This equation is motivated by the theoretical prediction that  that the distance along the jet of the emission zone goes as $r \propto \nu^{-1}$ when the jet is optically thick \citep{blandford/konigl:1979}. Eq.~\ref{eq:powlaw} follows because the phase shift in the KO model is due to light-travel times, and thus proportional to $r$. Fig.~\ref{plt:2131_phases} shows the best fit of Eq.~\ref{eq:powlaw} to the radio and mm data (Haystack, ALMA and ACT) for \pkstwo, with best-fitting parameters reported in Table~\ref{tab:phase_comparison}.\footnote{Including the infrared and optical frequencies result in a poor fit, so we omit them. This is not necessarily surprising: the emission is likely to be more complicated at these higher frequencies, which originate close to the central engine, thus affecting the scaling of the phase shifts.} Fits were performed with a Metropolis--Hastings Monte Carlo (MC) with flat priors in the ranges $b = (0.05, 4)$ and $a = (-2, -0.05)$. The priors were cut off before zero to prevent the MC from piling up near unphysical $a = 0$. The exact cutoff does not change the maximum likelihood; naturally, though, it alters the posterior credible regions that we use for our error bars, but this does not qualitatively alter our findings. To check the accuracy of the MC, we also performed fits with the Marquardt--Levenberg algorithm and found consistent best-fitting values. Note that we do not know the level of covariance between all of the data points, which are treated independently in our fits, but in reality there will be some degree of covariance (see Appendix~\ref{app:empirical_estimate}).

\begin{figure}[tb]
   \centering
   \footnotesize
   \begingroup
  \makeatletter
  \providecommand\color[2][]{\GenericError{(gnuplot) \space\space\space\@spaces}{Package color not loaded in conjunction with
      terminal option `colourtext'}{See the gnuplot documentation for explanation.}{Either use 'blacktext' in gnuplot or load the package
      color.sty in LaTeX.}\renewcommand\color[2][]{}}\providecommand\includegraphics[2][]{\GenericError{(gnuplot) \space\space\space\@spaces}{Package graphicx or graphics not loaded}{See the gnuplot documentation for explanation.}{The gnuplot epslatex terminal needs graphicx.sty or graphics.sty.}\renewcommand\includegraphics[2][]{}}\providecommand\rotatebox[2]{#2}\@ifundefined{ifGPcolor}{\newif\ifGPcolor
    \GPcolortrue
  }{}\@ifundefined{ifGPblacktext}{\newif\ifGPblacktext
    \GPblacktexttrue
  }{}\let\gplgaddtomacro\g@addto@macro
\gdef\gplbacktext{}\gdef\gplfronttext{}\makeatother
  \ifGPblacktext
\def\colorrgb#1{}\def\colorgray#1{}\else
\ifGPcolor
      \def\colorrgb#1{\color[rgb]{#1}}\def\colorgray#1{\color[gray]{#1}}\expandafter\def\csname LTw\endcsname{\color{white}}\expandafter\def\csname LTb\endcsname{\color{black}}\expandafter\def\csname LTa\endcsname{\color{black}}\expandafter\def\csname LT0\endcsname{\color[rgb]{1,0,0}}\expandafter\def\csname LT1\endcsname{\color[rgb]{0,1,0}}\expandafter\def\csname LT2\endcsname{\color[rgb]{0,0,1}}\expandafter\def\csname LT3\endcsname{\color[rgb]{1,0,1}}\expandafter\def\csname LT4\endcsname{\color[rgb]{0,1,1}}\expandafter\def\csname LT5\endcsname{\color[rgb]{1,1,0}}\expandafter\def\csname LT6\endcsname{\color[rgb]{0,0,0}}\expandafter\def\csname LT7\endcsname{\color[rgb]{1,0.3,0}}\expandafter\def\csname LT8\endcsname{\color[rgb]{0.5,0.5,0.5}}\else
\def\colorrgb#1{\color{black}}\def\colorgray#1{\color[gray]{#1}}\expandafter\def\csname LTw\endcsname{\color{white}}\expandafter\def\csname LTb\endcsname{\color{black}}\expandafter\def\csname LTa\endcsname{\color{black}}\expandafter\def\csname LT0\endcsname{\color{black}}\expandafter\def\csname LT1\endcsname{\color{black}}\expandafter\def\csname LT2\endcsname{\color{black}}\expandafter\def\csname LT3\endcsname{\color{black}}\expandafter\def\csname LT4\endcsname{\color{black}}\expandafter\def\csname LT5\endcsname{\color{black}}\expandafter\def\csname LT6\endcsname{\color{black}}\expandafter\def\csname LT7\endcsname{\color{black}}\expandafter\def\csname LT8\endcsname{\color{black}}\fi
  \fi
    \setlength{\unitlength}{0.0500bp}\ifx\gptboxheight\undefined \newlength{\gptboxheight}\newlength{\gptboxwidth}\newsavebox{\gptboxtext}\fi \setlength{\fboxrule}{0.5pt}\setlength{\fboxsep}{1pt}\definecolor{tbcol}{rgb}{1,1,1}\begin{picture}(5040.00,3600.00)\gplgaddtomacro\gplbacktext{\csname LTb\endcsname \put(734,966){\makebox(0,0)[r]{\strut{}$-0.2$}}\csname LTb\endcsname \put(734,1528){\makebox(0,0)[r]{\strut{}$-0.15$}}\csname LTb\endcsname \put(734,2090){\makebox(0,0)[r]{\strut{}$-0.1$}}\csname LTb\endcsname \put(734,2652){\makebox(0,0)[r]{\strut{}$-0.05$}}\csname LTb\endcsname \put(734,3214){\makebox(0,0)[r]{\strut{}$0$}}\csname LTb\endcsname \put(1089,432){\makebox(0,0){\strut{}$15$}}\csname LTb\endcsname \put(3220,432){\makebox(0,0){\strut{}$95$}}\csname LTb\endcsname \put(3724,432){\makebox(0,0){\strut{}$147$}}\csname LTb\endcsname \put(4216,432){\makebox(0,0){\strut{}$225$}}}\gplgaddtomacro\gplfronttext{\csname LTb\endcsname \put(1498,1395){\makebox(0,0)[l]{\strut{}\pkstwo (MCMC) }}\csname LTb\endcsname \put(1498,1199){\makebox(0,0)[l]{\strut{}\pkstwo (emp.)}}\csname LTb\endcsname \put(1498,1002){\makebox(0,0)[l]{\strut{}\pkszero (MCMC)}}\csname LTb\endcsname \put(1498,806){\makebox(0,0)[l]{\strut{}\pkszero (emp.)}}\csname LTb\endcsname \put(24,2006){\rotatebox{-270.00}{\makebox(0,0){\strut{}Phase Shift (cycles)}}}\csname LTb\endcsname \put(2779,137){\makebox(0,0){\strut{}Frequency (GHz)}}}\gplbacktext
    \put(0,0){\includegraphics[width={252.00bp},height={180.00bp}]{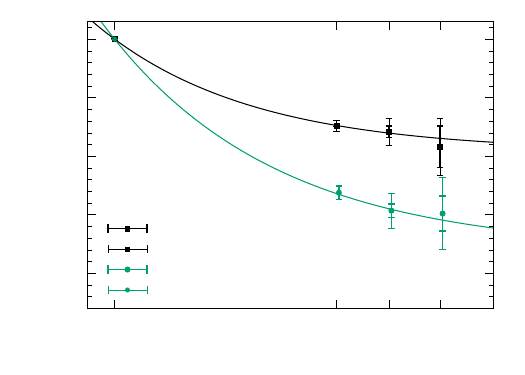}}\gplfronttext
  \end{picture}\endgroup
    \caption{Phase shifts of the ACT best-fit sinusoids relative to the OVRO 15\,GHz light curves for each of \pkstwo and \pkszero. The lines are the best fits to the power law of Eq.~\ref{eq:powlaw}, with only ACT data included. Best-fitting values are listed in Table~\ref{tab:phase_comparison}. Heavy errorbars are the fiducial uncertainties of the phase shifts from the MCMC fits (Sec.~\ref{sec:sinfit_method}) relative to 15\,GHz, while the light errorbars show an empirically-derived, relative phase uncertainty between 95\,GHz and 147/225\,GHz (see Appendix~\ref{app:empirical_estimate_backgnd}). Fits are done with the fiducial uncertainties. The \pkszero points have been shifted slightly to the right for ease of viewing.}
   \label{plt:phase_comparison}
\end{figure}

The best-fitting index of $b = -0.3\pm0.1$ for the \pkstwo mm data suggests that the jet is not optically thick in these emission zones. This may be compared to the finding in K25, based on the emission spectrum of \pkstwo, that the jet is optically thick below 70\,GHz and optically thin above. To explore whether our result is sensitive to particular combination of frequencies, we fit different ranges. First, since the 345\,GHz point from ALMA is a notable outlier to our fit---though the goodness of fit is acceptable, with PTE = 0.07---we excluded this point and obtained $b = -0.46^{+0.1}_{-0.09}$. (All fit results are shown in Table~\ref{tab:phase_comparison}.) This may suggest that the jet becomes optically thinner above 225\,GHz, but the difference in fits is marginal.
If we restrict our fits to the phase shifts of the ACT data alone (95--225\,GHz), the best-fitting index is $b = -0.8^{+0.1}_{-0.6}$, which has too broad a posterior to be informative; Fig.~\ref{plt:phase_mcmc} shows that the ACT data in isolation are not sufficient to break the degeneracy of the power law amplitude, $a$, and index, $b$. Thus, more phase-shift data are needed to probe whether the optical thickness changes through the radio, mm and submm emission regions, but our results are most consistent with the jet being optically thin in these zones.

Fig.~\ref{plt:phase_comparison} shows the frequency--phase relations from the ACT data of \pkstwo and \pkszero together with the best-fitting power laws. Interestingly, the best-fitting values for the power law index are similar between the two AGNs (0.8 and 0.7 for \pkstwo and \pkszero, respectively), both of which are consistent with $b = -1$. However, as mentioned above, the data support a wide range of $b$ values (see Fig.~\ref{plt:phase_mcmc}) so this similarity may be a coincidence.

In summary, two points should be noted about the phase shifts:
\begin{enumerate}
    \item For both SMBHB candidates, the phase shift between 15\,GHz and 95, 147 and 225\,GHz is highly significant: $\chi^2$ for the null hypothesis of no phase shift is 548 and 1288, with three degrees of freedom, for \pkstwo and \pkszero, respectively. Thus, regardless of the details of how the slope behaves in the radio and mm regime (point 2, below), the ACT data provide a strong detection of the phase shift between the radio and mm sinusoidal emission.
    
    \item When all radio and mm data are considered, the power law index is inconsistent with the jet of \pkstwo being optically thick. If only the ACT data are considered, \pkstwo and \pkszero have similar best-fitting values for the power law index, but support a wide range of values. More data are needed to understand if there is a change in jet properties in the mm zones.
\end{enumerate}

Improving on these results above would not only benefit from more data, but would also require better understanding of the uncertainties of the phase shifts. As discussed in K25, intrinsic variations in the jet that are superimposed on the sinusoidal variations can add covariance between the light curves at different frequencies, complicating the noise model used in fitting. We discuss this further in Appendix~\ref{app:empirical_estimate}, where we also make an empirical estimate of the phase shift uncertainties by injecting sine waves into other, real ACT light curves. These uncertainties are shown with light errorbars in Fig.~\ref{plt:phase_comparison}. Our analysis suggests that the MCMC-derived errorbars listed in Tables~\ref{tab:2131_fits} and \ref{tab:0805_fits} and shown with heavy errorbars in Fig.~\ref{plt:phase_comparison}, are possibly underestimated. However, as we argue in Appendix~\ref{app:empirical_estimate}, light curves of more SMBHB candidates are needed to confirm such an assessment.

\subsection{Achromatic variability}

Both the sinusoidal variation in emission (Sec.~\ref{sec:sinfit}) and the frequency-dependent phase shifts just discussed (Sec.~\ref{sec:phase_shifts}) provide important constraints on the physical origin of the emission and are predicted by the KO model. The model also predicts that the fractional amplitude of the sinusoidal variations should be achromatic, assuming the spectral index $\alpha$ is constant, since the rest-frame flux density, $S'$, gets boosted to $S = \mathcal{D}^{2-\alpha}S'$, where $\mathcal{D}$ is the Doppler factor (K25). This is a third feature we verify in our data. The fractional change in intensity, expressed as $A/\bar{S}$, is similar for all frequencies (see Tables~\ref{tab:2131_fits} and \ref{tab:0805_fits}). K25 noted that it only varied by a factor of $\sim$2--3 over five decades in frequency, though their picture is complicated somewhat by the fact that the data were taken at several epochs. In contrast, K25 show that the OVRO and ALMA data, which were contemporaneous, show remarkably achromatic behaviour. In the OVRO+ACT data at 15, 95, 147 and 225\,GHz, we find $A / \bar{S} = (21\pm2)\,\%$ for \pkstwo, where the quoted uncertainty is the standard deviation of the ratios at the four frequencies. As with the other SMBHB properties, we once again confirm for the first time that this expected property is found in a second SMBHB candidate: \pkszero exhibits achromaticity, with $A / \bar{S} = (43\pm2)\,\%$.

\section{Interpreting the observations}\label{sec:newmodsims}

\subsection{A modified Kinetic Orbital (MKO) model}\label{ssec:mko}

\begin{figure}[tb]
   \setlength{\unitlength}{1cm}
   \begin{picture}(9.0,3.5)
     \definecolor{mko_green}{HTML}{127622}
     \definecolor{mko_blue}{HTML}{2a6099}
     \definecolor{mko_grey}{HTML}{666666}
     \definecolor{mko_deepred}{HTML}{8d281e}
     \definecolor{mko_litered}{HTML}{ff6d6d}
     \put(0,0){\includegraphics[width=9.0cm]{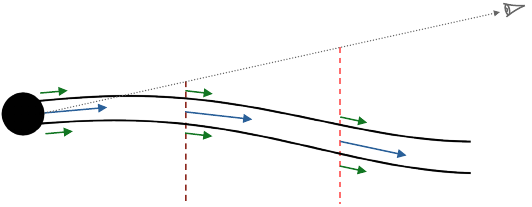}}
     \put(0.85,2.25){\makebox(0,0)[l]{\small$\color{mko_green}\vec{\beta}_{\mathrm{w}}$}}
     \put(1.85,1.65){\makebox(0,0)[l]{\small$\color{mko_blue}\vec{\beta}$}}
     \put(5.85,2.925){\rotatebox{12.5}{\makebox(0,0)[l]{\small\color{mko_grey}Line of sight}}}
     \put(3.3,0.25){\makebox(0,0)[l]{\small\color{mko_deepred}225\,GHz}}
     \put(5.85,0.25){\makebox(0,0)[l]{\small\color{mko_litered}15\,GHz}}
   \end{picture}
\caption{A 2D slice through the wind-collimated jet, illustrating the MKO model. The relativistic flow velocity, $\vec{\beta}$, at a particular point in the jet is parallel to the local wind velocity, $\vec{\beta}_w$. The wind propagates semi-ballistically, so different distances along the jet have local wind velocity directions out of phase. Since it is the aberration caused by the changing jet velocity that modulates the observed flux, frequencies originating from different locations in the jet that are observed simultaneously are at different phases in their sinusoidal flux variations.}
    \label{plt:mko_diagram}
\end{figure}

In this section we present a modified KO model, which we will refer to here and in future papers as the `MKO' model.  In the original KO model (\citealt{2017MNRAS.465..161S}; O22), the emitting region moves ballistically from one of the SMBH, producing a jet propagating at relativistic speeds and twisted into a conical helix by the orbital motion of the SMBH engine. The phase shift in the observed sinusoidal variations in flux between two frequencies, $\Delta\phi$, is due to the light-travel time between the two different zones of emission along the jet. Because the jet's velocity, $\beta = v / c$, is relativistic, in the observer's frame the jet material from the deeper zone travels almost as fast as the emitted light, such that its distance from the shallower zone, $\Delta r$, appears to be (K25):
\begin{equation}
\label{eq:distance}
    \Delta r = \beta(1+\beta) \gamma^2 c P_{\mathrm{B}} \Delta \phi \approx 2\gamma^2 c P_{\mathrm{B}} \Delta\phi,
\end{equation}
where $\gamma = (1-\beta)^{-1/2}$ is the Lorentz factor, $c$ is the speed of light, $P_{\mathrm{B}}$ is the binary orbital period (in the rest frame) and $\Delta\phi$ is in units of cycles. For high Lorentz factors, this implies that the deeper zone has wound through multiple helical rotations before the light is emitted from the shallower zone, even though the observed phase difference is a fraction of a cycle. For instance, K25 use VLBI measurements to obtain $\gamma = 10$ for \pkstwo; using our best-fitting period and phase differences (Table~\ref{tab:2131_phases}), this yields a distance of ${\sim}18$\,pc and ${\sim}26$ helical windings between the 15\,GHz and 95\,GHz emission zones. This is a potential problem with the simple KO model, because it is hard to explain how the jet can remain in a coherent helical pattern over such a distance. It would seem more likely that the phase would get smeared out.

In our new MKO model, the outflowing, ultrarelativistic jet containing the emitting electrons is not launched ballistically by its black hole source but instead follows a channel defined by a slower, mildly relativstic or sub-relativstic hydromagnetic wind that confines it (Sullivan et al. in prep). This semi-ballistic wind, possibly launched from the disc, confines the relativistic flow, dragging it around with the binary motion (see Fig.~\ref{plt:mko_diagram}). The channel may only be defined for a few helical windings, but because the wind has a low Lorentz factor (c.f., Eq.~\ref{eq:distance}) this suffices to account for sinusoidal variation due to the varying angle of the jet line of sight to the observer. The jet--wind complex forms a helix with pattern speed set by the wind rather than the relativistic flow. The different emission regions can thus come from a range of radii with each frequency's minimum radius separated by less than a wavelength of the conical helix $\lambda_h = \beta_w c P_\mathrm{B}$ and the average radius need not be separated by more than ${\sim}10\,\lambda_h$. 

Our analysis of the sinusoid phase shifts in Sec.~\ref{sec:phase_shifts} showed that the ACT data alone cannot constrain the power law $\Delta\phi \propto \nu^b$ well, but that when Haystack and ALMA data are included, there is evidence of $b < 1$ for \pkstwo. This differs from the classic scaling expected for the depth of a marginally optically thick photosphere as a function of frequency $r\propto\nu^{-1}$ \citep{blandford/konigl:1979} and would suggest that the emission does not come entirely from an optically thick photosphere as in the classic picture but instead may be optically thin. If optically thin, the observed emission at each radius would also contain the tail of the local synchrotron spectrum (Sullivan et al. in prep), and the frequency of the emissivity cut-off would thus decrease with distance along the jet.

\subsection{Simulations that address the intermittency of the periodicity}\label{ssec:bh_sims}

An observed feature of both \pkstwo and \pkszero that is not explained by the MKO model as presented thus far is the intermittency of the sinusoid pattern (see Sec.~\ref{sec:pks0805}). Here, we provide for the first time a possible physical explanation for this phenomenon. It is based upon the possibility of precession or disturbances in the SMBH accretion disc temporarily disrupting the production of a jet. Synchrotron emission from much further out in the jet is responsible for the baseline synchrotron emission (as found by O22 via VLBI measurements of \pkstwo), while the sinusoidal component is produced by the MKO mechanism near the central engine. When the jet is disrupted at its base, the sinusoidal signal shuts off, but when the jet relaunches the new sinusoid is in phase with the previous sinusoid because the SMBHB orbit has continued unabated.

This picture is supported by recent numerical simulations which have have extensively investigated black hole accretion flows; see, e.g., \citet{Davis:2020wea} for a recent review. Three-dimensional simulations of magnetised circumbinary accretion, in particular on black hole binaries, have also advanced considerably \citep{Farris:2012ux,Gold:2013zma,Manikantan:2024giq,Ennoggi:2025nht,Avara:2023ztw}. In the context of circumbinary disc formation from intermediate galactic scale accretion, \citet{wang2025galactic} have recently shown that realistic circumbinary discs might be strongly warped and tilted (\citealt{nixon2013tearing}; see also \citealt{Guo:2024gqc} for a similar study on single black holes). Although these simulations focus on equal-mass black hole binaries, the disc tilt originates from large-scale inflow and is expected to persist in systems with unequal mass ratios. These tilted accretion discs launch jets along the disc's rotation axis rather than the black hole's, as shown by \citet{liska2018formation}, who employed 3D general relativistic magnetohydrodynamic simulations. \citet{liska2018formation} showed how strong magnetic flux can partially realign the inner disc and jets with the black hole spin. On longer timescales, the entire disc-jet system undergoes Lense-Thirring precession \citep{lense1918einfluss}. This precession can produce curved jet morphologies and, in some cases, disrupt the accretion flow and shut off the jets. Misaligned circumbinary accretion discs around a binary black hole system lead to disc tearing, caused by strong differential torques \citep{nixon2013tearing, kaaz2025extreme}, which can result in jets shutting off. As such, realistic accretion environments might naturally be able to explain the temporary absence of a jet, with the timescale potentially being correlated with the inner disc precession timescale.

Even in the absence of disc tilt and precession, the accretion flow onto one constituent black hole in the binary can naturally feature variability. \citet{lalakos2025universal} have shown that extremely long-duration 3D general relativistic magnetohydrodynamic Bondi-like simulations \citep[see also][]{Cho:2023wqr,Galishnikova:2024nsk,Guo:2025sjb} can exhibit transitions between a magnetically arrested disc (MAD) state \citep{2003PASJ...55L..69N,2011MNRAS.418L..79T}, characterized by strong, stable jets, and a rocking accretion disc (RAD; \citealt{lalakos2024jets}). In the RAD state, the gas angular momentum is randomly oriented and chaotic, which leads to violently precessing inner accretion discs. The RAD jets are weaker as they tilt, wobble, and dissipate within a few thousand gravitational radii. These simulations show that MAD-to-RAD cycles occur on timescales of $\tau \sim 10^5 \, r_{\mathrm{g}}\,c^{-1}$, where $r_{\mathrm{g}}$ is the black hole's gravitational radius. For a $10^8\,\mathrm{M}_\odot$ black hole, $\tau \sim 600$\,days---comparable to the orbital period of our SMBHB candidates. In this scenario, continued monitoring of such systems may reveal multiple accretion and jet activity cycles tied to the Bondi accretion timescale of the individual black hole rather than the binary's orbital period.

\section{The value of millimetre light curves for SMBHB searches}\label{sec:cmparison}

\begin{figure*}[tb]
   \centering
\begingroup
  \makeatletter
  \providecommand\color[2][]{\GenericError{(gnuplot) \space\space\space\@spaces}{Package color not loaded in conjunction with
      terminal option `colourtext'}{See the gnuplot documentation for explanation.}{Either use 'blacktext' in gnuplot or load the package
      color.sty in LaTeX.}\renewcommand\color[2][]{}}\providecommand\includegraphics[2][]{\GenericError{(gnuplot) \space\space\space\@spaces}{Package graphicx or graphics not loaded}{See the gnuplot documentation for explanation.}{The gnuplot epslatex terminal needs graphicx.sty or graphics.sty.}\renewcommand\includegraphics[2][]{}}\providecommand\rotatebox[2]{#2}\@ifundefined{ifGPcolor}{\newif\ifGPcolor
    \GPcolortrue
  }{}\@ifundefined{ifGPblacktext}{\newif\ifGPblacktext
    \GPblacktexttrue
  }{}\let\gplgaddtomacro\g@addto@macro
\gdef\gplbacktext{}\gdef\gplfronttext{}\makeatother
  \ifGPblacktext
\def\colorrgb#1{}\def\colorgray#1{}\else
\ifGPcolor
      \def\colorrgb#1{\color[rgb]{#1}}\def\colorgray#1{\color[gray]{#1}}\expandafter\def\csname LTw\endcsname{\color{white}}\expandafter\def\csname LTb\endcsname{\color{black}}\expandafter\def\csname LTa\endcsname{\color{black}}\expandafter\def\csname LT0\endcsname{\color[rgb]{1,0,0}}\expandafter\def\csname LT1\endcsname{\color[rgb]{0,1,0}}\expandafter\def\csname LT2\endcsname{\color[rgb]{0,0,1}}\expandafter\def\csname LT3\endcsname{\color[rgb]{1,0,1}}\expandafter\def\csname LT4\endcsname{\color[rgb]{0,1,1}}\expandafter\def\csname LT5\endcsname{\color[rgb]{1,1,0}}\expandafter\def\csname LT6\endcsname{\color[rgb]{0,0,0}}\expandafter\def\csname LT7\endcsname{\color[rgb]{1,0.3,0}}\expandafter\def\csname LT8\endcsname{\color[rgb]{0.5,0.5,0.5}}\else
\def\colorrgb#1{\color{black}}\def\colorgray#1{\color[gray]{#1}}\expandafter\def\csname LTw\endcsname{\color{white}}\expandafter\def\csname LTb\endcsname{\color{black}}\expandafter\def\csname LTa\endcsname{\color{black}}\expandafter\def\csname LT0\endcsname{\color{black}}\expandafter\def\csname LT1\endcsname{\color{black}}\expandafter\def\csname LT2\endcsname{\color{black}}\expandafter\def\csname LT3\endcsname{\color{black}}\expandafter\def\csname LT4\endcsname{\color{black}}\expandafter\def\csname LT5\endcsname{\color{black}}\expandafter\def\csname LT6\endcsname{\color{black}}\expandafter\def\csname LT7\endcsname{\color{black}}\expandafter\def\csname LT8\endcsname{\color{black}}\fi
  \fi
    \setlength{\unitlength}{0.0500bp}\ifx\gptboxheight\undefined \newlength{\gptboxheight}\newlength{\gptboxwidth}\newsavebox{\gptboxtext}\fi \setlength{\fboxrule}{0.5pt}\setlength{\fboxsep}{1pt}\definecolor{tbcol}{rgb}{1,1,1}\begin{picture}(10080.00,4320.00)\gplgaddtomacro\gplbacktext{}\gplgaddtomacro\gplfronttext{\csname LTb\endcsname \put(2305,2754){\makebox(0,0)[r]{\strut{}OVRO 15\,GHz}}\csname LTb\endcsname \put(2305,2557){\makebox(0,0)[r]{\strut{}CRTS $V$-band}}\csname LTb\endcsname \put(2305,2361){\makebox(0,0)[r]{\strut{}ZTF $g$-band}}\csname LTb\endcsname \put(538,4103){\makebox(0,0)[r]{\strut{}15}}\csname LTb\endcsname \put(538,3669){\makebox(0,0)[r]{\strut{}15.5}}\csname LTb\endcsname \put(538,3234){\makebox(0,0)[r]{\strut{}16}}\csname LTb\endcsname \put(538,2800){\makebox(0,0)[r]{\strut{}16.5}}\csname LTb\endcsname \put(538,2366){\makebox(0,0)[r]{\strut{}17}}\csname LTb\endcsname \put(538,1931){\makebox(0,0)[r]{\strut{}17.5}}\csname LTb\endcsname \put(538,1497){\makebox(0,0)[r]{\strut{}18}}\csname LTb\endcsname \put(538,1063){\makebox(0,0)[r]{\strut{}18.5}}\csname LTb\endcsname \put(538,629){\makebox(0,0)[r]{\strut{}19}}\csname LTb\endcsname \put(1428,432){\makebox(0,0){\strut{}$54000$}}\csname LTb\endcsname \put(2749,432){\makebox(0,0){\strut{}$55000$}}\csname LTb\endcsname \put(4071,432){\makebox(0,0){\strut{}$56000$}}\csname LTb\endcsname \put(5392,432){\makebox(0,0){\strut{}$57000$}}\csname LTb\endcsname \put(6713,432){\makebox(0,0){\strut{}$58000$}}\csname LTb\endcsname \put(8034,432){\makebox(0,0){\strut{}$59000$}}\csname LTb\endcsname \put(9355,432){\makebox(0,0){\strut{}$60000$}}\csname LTb\endcsname \put(9717,2232){\makebox(0,0)[l]{\strut{}0}}\csname LTb\endcsname \put(9717,2767){\makebox(0,0)[l]{\strut{}1}}\csname LTb\endcsname \put(9717,3301){\makebox(0,0)[l]{\strut{}2}}\csname LTb\endcsname \put(9717,3836){\makebox(0,0)[l]{\strut{}3}}\csname LTb\endcsname \put(1080,4201){\makebox(0,0){\strut{}2006}}\csname LTb\endcsname \put(2044,4201){\makebox(0,0){\strut{}2008}}\csname LTb\endcsname \put(3010,4201){\makebox(0,0){\strut{}2010}}\csname LTb\endcsname \put(3974,4201){\makebox(0,0){\strut{}2012}}\csname LTb\endcsname \put(4940,4201){\makebox(0,0){\strut{}2014}}\csname LTb\endcsname \put(5904,4201){\makebox(0,0){\strut{}2016}}\csname LTb\endcsname \put(6870,4201){\makebox(0,0){\strut{}2018}}\csname LTb\endcsname \put(7834,4201){\makebox(0,0){\strut{}2020}}\csname LTb\endcsname \put(8800,4201){\makebox(0,0){\strut{}2022}}\csname LTb\endcsname \put(73,2366){\rotatebox{-270.00}{\makebox(0,0){\strut{}Optical Magnitude}}}\csname LTb\endcsname \put(9962,2366){\rotatebox{-270.00}{\makebox(0,0){\strut{}Flux Density (Jy)}}}\csname LTb\endcsname \put(5127,137){\makebox(0,0){\strut{}Modified Julian Day}}}\gplbacktext
    \put(0,0){\includegraphics[width={504.00bp},height={216.00bp}]{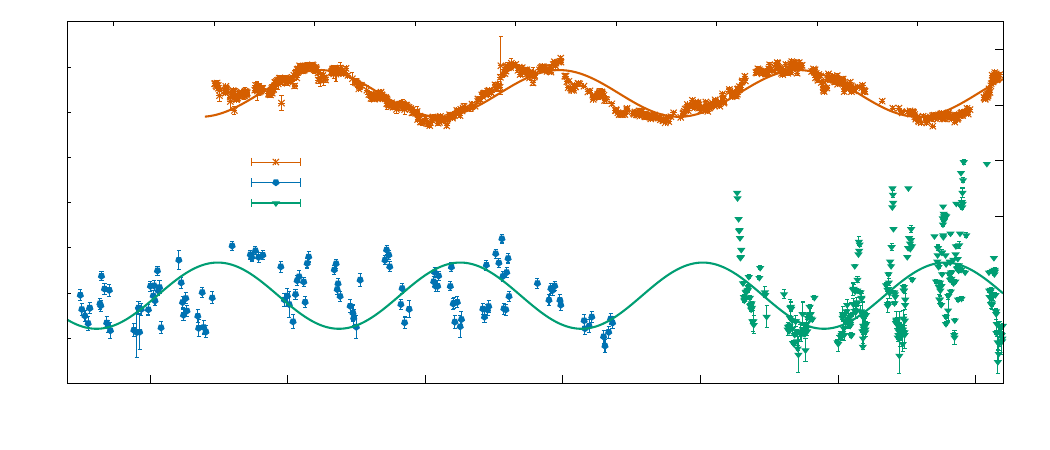}}\gplfronttext
  \end{picture}\endgroup
       \caption{Comparison of radio and optical light curves of \pkstwo. The sine waves shown in the figure come from tables~1 and 4 of K25 for the radio and ZTF optical data, respectively. The fit to the ZTF data has been extrapolated back in time to show that the cadence of the CRTS is not sufficient to easily discover this sinusoidal pattern. See text for additional details.}
         \label{plt:optrad}
\end{figure*}

A notable feature of the mm light curves shown in Figs.~\ref{plt:2131} and \ref{plt:0805} is their short-term deviations from the sinusoids. However, for the 95 and 147\,GHz light curves, the sinusoidal variation dominates by a factor of $\sim$2--3.5 (see the \snrsin values in Table~\ref{tab:smbhb_props}). Thus, although the mm light curves do not have the same sensitivity as the OVRO 15\,GHz light curves, and hence the uncertainties on individual observations are much larger in the ACT than in the OVRO data, this does not mean that ACT is less sensitive to sinusoidal light curves in blazars than the OVRO, provided we consider equal timespans.\footnote{The 225\,GHz channel has more noise due to the larger atmospheric contamination it suffers, and also has lower fluxes due to the falling synchrotron spectrum. It will therefore not be as important a channel for SMBHB detection. Nonetheless, for our SMBHB candidates, its SNR of $\approx 1$ (Table~\ref{tab:act_phase_uncertainty}), will still allow 225\,GHz data to be used to study phase shifts and spectra.} This argues strongly for continued, long term observations with mm survey instruments, such as the South Pole Telescope \citep{carlstrom/etal:2011,benson/etal:2014}, the Simons Observatory (SO; \citealt{so_collaboration:2019,so_collaboration:2025}) and the CCAT Observatory \citep{ccat:2023} for multi-messenger astronomy.  

The promise of cm, mm, and submm data for identifying and studying SMBHB is evident when they are placed side by side with optical data. Fig.~\ref{plt:optrad} shows the OVRO 15\,GHz light curve of \pkstwo together with optical light curves of the same object from CRTS and ZTF. As discussed in K25, the ZTF data show much stronger short-term variations than the radio--submm data, but when averaged over a year they follow a sine wave well and have a period consistent with that derived from the OVRO and Haystack 15\,GHz observations. To further explore this point, in Appendix \ref{app:sim} we give a detailed account of statistical tests we have carried out on the CRTS and ZTF data for \pkstwo and show that the optical data alone do not provide compelling evidence for the periodicity we see in \pkstwo, unless they are sampled about 3 times a week and then averaged for a long period ($\sim$~several months) to eliminate the rapid large fluctuations at optical wavelengths. This is illustrated in Fig.~\ref{plt:optrad} by the sine wave, fitted to ZTF data by K25 (reported in their table~4), and here extrapolated back to the period covered by the CRTS. There is no clear correspondence between the curve and the CRTS data points, suggesting that searches for SMBHB candidates similar to \pkstwo are unlikely to be found in the CRTS because the cadence is not frequent enough. The ZTF cadence is two days, and that clearly works well, although the short-term variations are much greater than at cm--mm--submm wavelengths. We do not offer any specific theories here about the origin of this behaviour, but it is reasonable to suppose that activity near the central engine could produce very different short-term variations than those seen from the more distant zones producing the radio and mm radiation.

 \citet{ren/etal:2025} also found only weak evidence for periodicity from a Lomb-Scargle analysis of infrared (\textit{WISE}, \textit{NEOWISE}) and optical (ZTF, CRTS, ATLAS) light curves of \pkstwo, further highlighting the importance of cm-mm-submm data for this class of SMBHB candidate.

The cm, mm, and submm observations also have the great advantage that they can be sampled year-round, while at optical frequencies the daylight sky is bright compared to even the brightest optical blazar, 3C~273, which has an apparent optical magnitude of $V \sim 13$,\footnote{See the NASA/IPAC Extragalactic Database (NED).} so year-round optical monitoring of blazars is not possible.\footnote{While AGN can be observed in daylight in the mm, it is still the case that mid-latitude CMB surveys will have observing gaps for most AGNs for part of the year---about one month for SO---due to Sun- and Moon-avoidance in their observation strategies \citep{so_collaboration:2025}.\label{fn:sun_avoidance}} Nevertheless, there is evidence for SMBHB candidates that show more prominent sinusoidal variations at optical wavelengths.  PG~1302$-$102 might be just such an example \citep{2015Natur.518...74G}: this source also shows radio periodicity at the same period as detected in the optical \citep{2018A&A...615A.123Q}. Recently, \citet{huijse/etal:2025} reported the detection of 181 SMBHB candidates from a search of 770,100 AGN that were monitored for 34 months in the optical by the \textit{Gaia} satellite. It is therefore likely that optical and cm--mm--submm observations will be highly synergistic for discovering and characterising SMBHBs.

\section{Summary and discussion}\label{sec:discussion}

ACT mm data at 95, 147 and 225\,GHz have confirmed the presence of a sinusoidal variation in the light curves of \pkstwo and \pkszero that had been detected with high significance in OVRO 15\,GHz data by O22, D25 and K25. Our main findings are as follows:

\begin{enumerate}
    \item The ACT 95\,GHz light curve of \pkstwo is in good agreement with the ALMA 91.5 and 103.5\,GHz light curves, including in their small variations superimposed on the larger sinusoid variation. This confirms that the ACT calibration and data reduction of light curves are working well.
    
    \item ACT light curves of \pkszero, for which no ALMA data are available, corroborate the finding in D25 that this source is a SMBHB candidate. Especially compelling is the fact that the sinusoid appears to turn off in the ACT light curves in the same manner as in the OVRO data in mid-2020 (see Fig.~\ref{plt:0805}). This provides further evidence that: (a) the long time-scale variations of the mm light curves are tracing the same underlying physics as the radio light curve and (b) that the phenomenon of the sinusoid stopping and starting, discovered in O22 for \pkstwo, is likely a common feature of SMBHB candidates. It can be explained as the temporary disruption of the jet by precession or by MAD-to-RAD transitions of the accretion disc. Further data will help clarify these hypotheses, and our results show that mm observations can play a key role in detecting this behaviour.

    \item The ACT data confirm that the sinusoidal variations at higher frequencies are shifted in phase relative to radio frequencies, a feature that is naturally explained in the MKO model by light-travel times due to higher frequencies originating from deeper in the jet. In both cases the mm phase leads the radio phase in time.

    \item A power-law fit to the phase shifts in \pkstwo in the radio to submm-regime suggest that the jet in these emission zones is optically thin, but it is unclear whether the jet properties are uniform throughout the regions. ACT data alone cannot well-constrain the power-law index for either \pkszero or \pkstwo given current uncertainties.

    \item The sinusoidal signal is achromatic from 15 to 225\,GHz, i.e., the ratio of its amplitude to offset, $A/\bar{S}$, is nearly constant at all measured frequencies in this range. This is expected under the MKO model and is further evidence for these objects being SMBHBs.

    \item In \pkstwo, the ratio of the sinusoid amplitude to the non-sinusoidal variations in the mm data is significantly higher than in optical data, which are contaminated by large, short term variability from the point of view of detecting SMBHB candidates. It seems likely that some SMBHB candidates will appear more cleanly in the optical (e.g., PG~1302$-$102, \citealt{2015Natur.518...74G}; see also \citealt{huijse/etal:2025}), but our findings provide strong motivation for using mm surveys to search for SMBHB. Such searches will be complementary to radio, submm and optical searches.
\end{enumerate}

In addition to these empirical findings, we have presented the MKO model which, when combined with recent findings from simulated BH accretion discs, explains all four of the properties observed in the two SMBHB candidates: (a) the flux density exhibits a sinusoidal pattern; (b) the sinusoid is monochromatic, in that it has the same period and a similar amplitude to mean-flux ratio over a broad range of frequencies; (c) the phase of the sinusoid shifts monotonically with frequency; (d) the sinusoidal variations are intermittent on time scales $\gtrsim$ the SMBHB period.

Our results demonstrate the value of high cadence, wide area surveys of the mm and submm sky available from CMB and mm/sub-mm survey experiments such as SPT, SO and CCAT. In particular, the large aperture telescope (LAT) of SO will observe $25\,000\,\mathrm{deg}^2$ of the sky from 2025 to 2034 at 27, 39, 93, 145, 225 and 280\,GHz with a planned cadence of 1--2 days \citep{so_collaboration:2025}.\footnote{Though see footnote~\ref{fn:sun_avoidance}.} We use their goal sensitivity for single observations at 93 GHz to forecast the number of AGN that SO can monitor over a $21\,000\,\mathrm{deg}^2$ sky area (to exclude the Galactic plane), using the 90\,GHz radio number counts of \citet{lagache/etal:2020}. We find that SO will be able to monitor ${\sim}8000$ AGN with a signal-to-noise ratio > 5, if we allow six measurements to be binned together (i.e., corresponding to a 1--2 week cadence). The number will increase to ${\sim}12\,000$ when the Advanced SO upgrades, which will double the number of detectors, are completed by 2028. Estimating the rates of blazars in SMBHB depends on a number of factors, as discussed in K25 and D25, but these authors conclude that with current information, ${\sim}1$ in 100 blazars may be in SMBHB systems. If this is correct then SO should be able to observe $\mathcal{O}(100)$ SMBHB candidates. This large sample size will better constrain the SMBHB occurrence rate.
At this point \pkstwo is the only blazar in which sinusoidal variations have been observed at both optical and radio wavelengths,\footnote{The report of 181 SMBHB candidates found in optical \textit{Gaia} data by \citeauthor{huijse/etal:2025}~(\citeyear{huijse/etal:2025}; see Sec.~\ref{sec:cmparison}) was posted to the arXiv while this paper was in review; we have not yet followed up these candidates in our data.} but this is unlikely to be an isolated case given the broadband sinusoidal variations of \pkszero that we present in this paper, making a compelling case for joint searches for SMBHB candidates spanning from the radio to optical.

The nature of sinusoidal blazar signals and their potential connection with supermassive black hole binaries is of fundamental interest for both astrophysics and cosmology. Multi-wavelength information about more systems and over more periods, as will be possible with upcoming mm-wave surveys, will establish the correct model of emission and its connection to binary properties, including the duty cycle for sinusoidal emission and the selection function for observing binaries. Significant detections of correlated gravitational wave and electromagnetic signals from particular SMBHB systems are also imminent \citep{2025arXiv250816534A}. In principle, the absolute phase of mm-discovered SMBHB could be obtained from optical light curves, since the optical emission is believed to originate close to the AGN core and would therefore be virtually in phase with the SMBHB orbit. Further, if velocities were obtained from optical spectra, the masses could be derived, and the expected gravitational wave phase and amplitude of each source could be determined. One could then mount coherent gravitational wave searches with millsecond pulsar timing arrays, rather than searching within the stochastic noise, potentially expanding their reach to weaker sources of gravitational waves. In turn, these observations and binary blazar models could provide more reliable observation-based estimates of the total binary population with orbital periods of years. This population is of intense interest for its production of gravitational waves: the Laser Interferometer Space Antenna (LISA) will directly measure the supermassive binary merger rate throughout the Hubble volume, and comparison with the total years-period binary population will constrain the complex galactic dynamics and astrophysical mechanisms driving binary mergers (e.g., \citealt{2015ApJ...812...72A,2020MNRAS.497..739N,2003ApJ...596..860M}). 

In addition, the close-orbit SMBHB population will likely provide a dominant contribution to the stochastic background of gravitational waves at nanohertz frequencies for pulsar timing arrays 
or astrometric probes \citep{2023ApJ...952L..37A}. This stochastic background complicates the prospects for detecting other cosmological sources of gravitational waves from the early universe \citep{2021PhRvD.103j3529B}, such as phase transitions or reheating after inflation (see \citealt{2023ApJ...951L..11A} for a recent overview). A detailed understanding of the SMBHB population enabled by multi-wavelength blazar observations may eventually be key to exploiting unique gravitational wave information about physics at the highest energy scales. 

\section*{Acknowledgements}

We thank an anonymous referee for comments that strengthened our analysis and prompted us to introduce the MKO model.

This work was supported by the U.S. National Science Foundation through awards AST-0408698, AST-0965625, and AST-1440226 for the ACT project, as well as awards PHY-0355328, PHY-0855887, and PHY-1214379. Funding was also provided by Princeton University, the University of Pennsylvania, and a Canada Foundation for Innovation (CFI) award to UBC.

ACT operated in the Parque Astron\'omico Atacama in northern Chile under the auspices of the Agencia Nacional de Investigaci\'on y Desarrollo (ANID; formerly Comisi\'on Nacional de Investigaci\'on Cient\'ifica y Tecnol\'ogica de Chile, or CONICYT). We thank the Republic of Chile for hosting ACT in the northern Atacama, and the local indigenous Licanantay communities whom we follow in observing and learning from the night sky.

The development of multichroic detectors and lenses for ACT was supported by NASA grants NNX13AE56G and NNX14AB58G. Detector research at NIST was supported by the NIST Innovations in Measurement Science program. Computing for ACT was performed using the Princeton Research Computing resources at Princeton University, the National Energy Research Scientific Computing Center (NERSC), and the Niagara supercomputer at the SciNet HPC Consortium. SciNet is funded by the CFI under the auspices of Compute Canada, the Government of Ontario, the Ontario Research Fund--Research Excellence, and the University of Toronto.

Colleagues at AstroNorte and RadioSky provided logistical support for ACT and kept operations in Chile running smoothly. We also thank the Mishrahi Fund and the Wilkinson Fund for their generous support of the project.

The OVRO programme was supported by NASA grants \hbox{NNG06GG1G}, \hbox{NNX08AW31G}, \hbox{NNX11A043G}, and \hbox{NNX13AQ89G} from~2006 to~2016 and NSF grants AST-0808050 and AST-1109911 from~2008 to~2014. This work is currently supported by NSF grants AST2407603 and AST2407604.

This is not an official SO Collaboration paper.

ADH acknowledges support from the Sutton Family Chair in Science, Christianity and Cultures, from the Faculty of Arts and Science, University of Toronto, and from the Natural Sciences and Engineering Research Council of Canada (NSERC) [RGPIN-2023-05014, DGECR-2023-00180], and is further grateful to the Lumen Christi Institute for support that enabled a research visit to the University of Chicago, during which significant portions of this work were done.
S.K. was funded by the European Union ERC-2022-STG -- BOOTES -- 101076343. Views and opinions expressed are however those of the author(s) only and do not necessarily reflect those of the European Union or the European Research Council Executive Agency. Neither the European Union nor the granting authority can be held responsible for them.
CS acknowledges support from the Agencia Nacional de Investigaci\'on y Desarrollo (ANID) through Basal project FB210003.
MJG acknowledges support from NSF grant AST-2108402.

This paper makes use of the following ALMA data: ADS/JAO.ALMA\#2011.0.00001.CAL. ALMA is a partnership of ESO (representing its member states), NSF (USA) and NINS (Japan), together with NRC (Canada), NSTC and ASIAA (Taiwan), and KASI (Republic of Korea), in cooperation with the Republic of Chile. The Joint ALMA Observatory is operated by ESO, AUI/NRAO and NAOJ. The National Radio Astronomy Observatory is a facility of the National Science Foundation operated under cooperative agreement by Associated Universities, Inc.

The CSS survey is funded by the National Aeronautics and Space Administration under Grant No. NNG05GF22G issued through the Science Mission Directorate Near-Earth Objects Observations Program. The CRTS survey is supported by the U.S.~National Science Foundation under grants AST-0909182 and AST-1313422.

ZTF is supported by the NSF under Grants No. AST-1440341 and AST-2034437 and a collaboration including current partners Caltech, IPAC, the Oskar Klein Center at Stockholm University, the University of Maryland, University of California, Berkeley , the University of Wisconsin at Milwaukee, University of Warwick, Ruhr University, Cornell University, Northwestern University and Drexel University. Operations are conducted by COO, IPAC, and UW.

This research has made use of the NASA/IPAC Extragalactic Database (NED), which is funded by the National Aeronautics and Space Administration and operated by the California Institute of Technology.

Some of the results/plots in this paper have been derived/produced using the following software: 
\textsc{Astropy},\footnote{http://www.astropy.org} a community-developed core Python package for Astronomy \citep{software:astropy/a,software:astropy/b,software:astropy/c};
\textsc{emcee} \citep{foreman-mackey/etal:2013};
\textsc{fastKDE} \citep{travis/etal:2014,travis/etal:2016};
\textsc{gnuplot};\footnote{http://www.gnuplot.info/}
\textsc{NumPy} \citep{software:numpy}; 
\textsc{pixell};\footnote{https://github.com/simonsobs/pixell} 
and \textsc{SciPy} \citep{software:scipy}.

\bibliographystyle{aa.bst}
\bibliography{references.bib}

\begin{appendix}

\section{ALMA sine-wave fit results}

K25 performed sine-wave fits to the OVRO and ALMA data for \pkstwo using the same methodology described in Sec.~\ref{sec:sinfit_method}, but only reported the best-fitting phase shifts. For completeness we include all the best-fitting parameters from this analysis in Table~\ref{tab:alma_2131_fits}.

\begin{table*}
\centering
\caption{Sine-wave fit results for \pkstwo.}
\begin{tabular}{lrrrr}
\hline \hline
Parameter & OVRO 15\,GHz & ALMA 91.5\,GHz & ALMA 103.5\,GHz & ALMA 345\,GHz \\
\hline
$P$ (days)            & \multicolumn{4}{c}{$2003\pm19$ (fixed)} \\
$A$   (Jy)         & $0.4385 \pm 0.0059$ & $0.3170 \pm 0.0066$  & $0.2995 \pm 0.0067$  & $0.1386 \pm 0.0060$ \\
$\phi_0$ (rad)     & $3.725 \pm 0.016$   & $3.268 \pm 0.027$    & $3.242 \pm 0.028$    & $2.886 \pm 0.054$ \\
$\bar{S}$ (Jy)         & $2.2022 \pm 0.0045$ & $1.4227 \pm 0.0054$  & $1.3536 \pm 0.0054$  & $0.7194 \pm 0.0044$ \\
$\xi$ (Jy)    & $0.0696 \pm 0.0035$ & $0.0731 \pm 0.0043$  & $0.0717 \pm 0.0042$  & $0.0466 \pm 0.0044$ \\
$\Delta\phi_0$ (cycles) & $0$                 & $-0.0727 \pm 0.0050$ & $-0.0769 \pm 0.0052$ & $-0.1336 \pm 0.0090$ \\
$\Delta\phi_0$ (days) & $0$                 & $-145.6 \pm 9.9$     & $-154 \pm 10$    & $-268 \pm 18$ \\
\hline
\end{tabular}
\tablefoot{The period, $P$, was determined from an initial fit including only the OVRO data in the range $57779 < \mathrm{MJD} < 60053$. The joint fit including OVRO and the three ALMA light curves kept the period fixed at this best fitting value. The fit uncertainties do not account for systematics due to superimposed, shorter term variations (see Appendix~\ref{app:empirical_estimate_backgnd}). The reference time for the phase, in MJD, is $t_0 = 59\,000$. The phase shift, $\Delta\phi_0$, is defined in Eq.~\ref{eq:phase_shift}.}
\label{tab:alma_2131_fits}
\end{table*}

\section{Empirical estimates of phase shift uncertainties}\label{app:empirical_estimate}

\subsection{Background}\label{app:empirical_estimate_backgnd}

Given the important information about SMBHB jet physics encoded in the phase of the sinusoidal variation of the light curves as a function of frequency, it is important to assess how well we can measure those phase shifts. There are at least two types of uncertainty associated with them. The \textit{absolute error} encodes how uncertain the phase measurement of a light curve is with respect to the true, underlying sinusoid. The \textit{relative error} encodes how uncertain the phase shifts at two different frequencies are with respect to each other, rather than to the true, underlying sinusoid. This is the more relevant uncertainty for probing how the location of the emission zone in the jet varies with frequency. If the light curves consisted of a pure sine wave with white noise, then the absolute and relative errors would be identical. In reality, the noise is correlated within single light curves and between light curves of different frequencies, and so the two types of uncertainty will be different.

Some background on physical origin of the light curves is in order. The radio flux density variations in blazars originate from a number of sites along the jet, but are dominated by the variations in the core, which is unresolved even by very long baseline imaging (VLBI; \citealt{Blandford2019}). These variations in blazars are believed to generally follow a power law spectrum (see, e.g., \citealt{Maxmoerbeck2014}), so it is to be expected that there will be ongoing, non-sinusoidal variations on both longer and shorter timescales than the sinusoidal variations. When a limited number of sinusoid cycles are measured, the superimposed variations can bias the phase shift. This is evident, for instance, in the 15\,GHz OVRO light curve of \pkstwo in Fig.~\ref{plt:2131}. One can see by eye that if the dashed orange curve, representing the best-fit sine wave to the OVRO data during the period when ALMA data are available, were continued half a cycle to the left, such that it extended into ${\sim}$2010--2014, it would not line up perfectly with the data points due to superimposed variations during this time range. Another feature to note in this plot is the fluctuation of the ALMA and ACT data points above the sine curve in the year 2018: one sees the \textit{same} trend at all the mm frequencies (Fig.~\ref{plt:J2134_alma-act_corr}), which also appears to be present in the 15\,GHz OVRO data earlier, at 2017.5 (Fig.~\ref{plt:2131}). In other words, random, non-sinusoidal fluctuations can be coherent across frequency bands. All these variations contribute to the uncertainty of the absolute phase measurement, and will be larger the smaller the number of cycles.

K25 distinguish two important types of variations, which we summarise here. `Type-A variations' are common across different observing frequencies, only shifted by the same amount as the sinusoid phase shift; an example was described at the end of the previous paragraph. `Type-B' variations, on the other hand, are unique at each frequency, or show a different frequency dependence from that of the sinusoid. If the light curves are dominated by Type-A variations, which could arise from variable fuelling of the jet, then the \textit{relative error} of the phase difference between frequencies will be considerably smaller than the absolute error, since all frequencies will have have the same bias in the absolute phase determination; if, on the other hand, they are dominated by Type-B variations, which could be caused by local disturbances along the jet, then the uncertainty of the relative phase difference would be commensurate with the uncertainty of the absolute phase.

Whether Type-A or Type-B variations dominate is an empirical question, and K25 conclude that they cannot make a firm estimate given the lack of a large sample of multi-frequency light curves of SMBHB candidates. Thus, our fitting procedure in Sec.~\ref{sec:sinfit_method}, while it includes a parameter for correlated noise within a single light curve, does not assign any noise covariance between different light curves; they are essentially independent fits, all using the same period that is measured in the OVRO data. As K25 point out, if all of the variations are Type-A, then any bias on the fits will be the same at all frequencies, and the independent fits will provide a good uncertainty estimate.

\begin{table}[tb]
  \centering
  \caption{Properties of SMBHB candidate ACT light curves}
  \begin{tabular}{|c|c|ccc|}
    \hline\hline
    AGN & Freq. & span\tablefootmark{a} & $\snrsin$ & $\snrvar$ \\
    & (GHz) & (days) & & \\
    \hline
    \pkstwo                                    &  95 & 2084 & 3.4  & 3.8 \\
    {\footnotesize\textit{period: 1931\,d}}  & 147 & 2197 & 2.5  & 2.9 \\
                                               & 225 & 1837 & 1.1  & 2.0 \\
    \hline
    \pkszero                                   &  95 & 1695 & 3.0  & 2.0 \\
    {\footnotesize\textit{period: 1402\,d}}    & 147 & 1708 & 2.2  & 1.6 \\
                                               & 225 & 1447 & 1.1  & 1.1 \\
    \hline
  \end{tabular}
  \label{tab:smbhb_props}
  \tablefoot{$\snrsin$ is the amplitude of the sine relative to its residuals (Eq.~\ref{eq:snrsin}, and $\snrvar$ is the rms of the residuals relative to the average flux uncertainty (Eq.~\ref{eq:snrvar}.)\\
    \tablefoottext{a}{The light curve length, ignoring gaps.}
  }
\end{table}

\subsection{Method}

\begin{figure}[tb]
   \centering
   \footnotesize
   \begingroup
  \makeatletter
  \providecommand\color[2][]{\GenericError{(gnuplot) \space\space\space\@spaces}{Package color not loaded in conjunction with
      terminal option `colourtext'}{See the gnuplot documentation for explanation.}{Either use 'blacktext' in gnuplot or load the package
      color.sty in LaTeX.}\renewcommand\color[2][]{}}\providecommand\includegraphics[2][]{\GenericError{(gnuplot) \space\space\space\@spaces}{Package graphicx or graphics not loaded}{See the gnuplot documentation for explanation.}{The gnuplot epslatex terminal needs graphicx.sty or graphics.sty.}\renewcommand\includegraphics[2][]{}}\providecommand\rotatebox[2]{#2}\@ifundefined{ifGPcolor}{\newif\ifGPcolor
    \GPcolortrue
  }{}\@ifundefined{ifGPblacktext}{\newif\ifGPblacktext
    \GPblacktexttrue
  }{}\let\gplgaddtomacro\g@addto@macro
\gdef\gplbacktext{}\gdef\gplfronttext{}\makeatother
  \ifGPblacktext
\def\colorrgb#1{}\def\colorgray#1{}\else
\ifGPcolor
      \def\colorrgb#1{\color[rgb]{#1}}\def\colorgray#1{\color[gray]{#1}}\expandafter\def\csname LTw\endcsname{\color{white}}\expandafter\def\csname LTb\endcsname{\color{black}}\expandafter\def\csname LTa\endcsname{\color{black}}\expandafter\def\csname LT0\endcsname{\color[rgb]{1,0,0}}\expandafter\def\csname LT1\endcsname{\color[rgb]{0,1,0}}\expandafter\def\csname LT2\endcsname{\color[rgb]{0,0,1}}\expandafter\def\csname LT3\endcsname{\color[rgb]{1,0,1}}\expandafter\def\csname LT4\endcsname{\color[rgb]{0,1,1}}\expandafter\def\csname LT5\endcsname{\color[rgb]{1,1,0}}\expandafter\def\csname LT6\endcsname{\color[rgb]{0,0,0}}\expandafter\def\csname LT7\endcsname{\color[rgb]{1,0.3,0}}\expandafter\def\csname LT8\endcsname{\color[rgb]{0.5,0.5,0.5}}\else
\def\colorrgb#1{\color{black}}\def\colorgray#1{\color[gray]{#1}}\expandafter\def\csname LTw\endcsname{\color{white}}\expandafter\def\csname LTb\endcsname{\color{black}}\expandafter\def\csname LTa\endcsname{\color{black}}\expandafter\def\csname LT0\endcsname{\color{black}}\expandafter\def\csname LT1\endcsname{\color{black}}\expandafter\def\csname LT2\endcsname{\color{black}}\expandafter\def\csname LT3\endcsname{\color{black}}\expandafter\def\csname LT4\endcsname{\color{black}}\expandafter\def\csname LT5\endcsname{\color{black}}\expandafter\def\csname LT6\endcsname{\color{black}}\expandafter\def\csname LT7\endcsname{\color{black}}\expandafter\def\csname LT8\endcsname{\color{black}}\fi
  \fi
    \setlength{\unitlength}{0.0500bp}\ifx\gptboxheight\undefined \newlength{\gptboxheight}\newlength{\gptboxwidth}\newsavebox{\gptboxtext}\fi \setlength{\fboxrule}{0.5pt}\setlength{\fboxsep}{1pt}\definecolor{tbcol}{rgb}{1,1,1}\begin{picture}(5040.00,5760.00)\gplgaddtomacro\gplbacktext{\csname LTb\endcsname \put(146,3673){\makebox(0,0)[r]{\strut{}}}\csname LTb\endcsname \put(146,3957){\makebox(0,0)[r]{\strut{}}}\csname LTb\endcsname \put(146,4241){\makebox(0,0)[r]{\strut{}}}\csname LTb\endcsname \put(146,4525){\makebox(0,0)[r]{\strut{}}}\csname LTb\endcsname \put(146,4809){\makebox(0,0)[r]{\strut{}}}\csname LTb\endcsname \put(146,5093){\makebox(0,0)[r]{\strut{}}}\csname LTb\endcsname \put(473,3477){\makebox(0,0){\strut{}}}\csname LTb\endcsname \put(931,3477){\makebox(0,0){\strut{}}}\csname LTb\endcsname \put(1388,3477){\makebox(0,0){\strut{}}}\csname LTb\endcsname \put(1846,3477){\makebox(0,0){\strut{}}}\csname LTb\endcsname \put(2304,3477){\makebox(0,0){\strut{}}}}\gplgaddtomacro\gplfronttext{\csname LTb\endcsname \put(1377,5297){\makebox(0,0){\strut{}PKS~2131$-$021}}}\gplgaddtomacro\gplbacktext{\csname LTb\endcsname \put(2412,3673){\makebox(0,0)[r]{\strut{}}}\csname LTb\endcsname \put(2412,3957){\makebox(0,0)[r]{\strut{}}}\csname LTb\endcsname \put(2412,4241){\makebox(0,0)[r]{\strut{}}}\csname LTb\endcsname \put(2412,4525){\makebox(0,0)[r]{\strut{}}}\csname LTb\endcsname \put(2412,4809){\makebox(0,0)[r]{\strut{}}}\csname LTb\endcsname \put(2412,5093){\makebox(0,0)[r]{\strut{}}}\csname LTb\endcsname \put(2733,3477){\makebox(0,0){\strut{}}}\csname LTb\endcsname \put(3181,3477){\makebox(0,0){\strut{}}}\csname LTb\endcsname \put(3629,3477){\makebox(0,0){\strut{}}}\csname LTb\endcsname \put(4077,3477){\makebox(0,0){\strut{}}}\csname LTb\endcsname \put(4524,3477){\makebox(0,0){\strut{}}}}\gplgaddtomacro\gplfronttext{\csname LTb\endcsname \put(4280,4973){\makebox(0,0)[r]{\strut{}f090 Abs. Err}}\csname LTb\endcsname \put(3618,5297){\makebox(0,0){\strut{}PKS~J0805$-$0111}}}\gplgaddtomacro\gplbacktext{\csname LTb\endcsname \put(146,2123){\makebox(0,0)[r]{\strut{}}}\csname LTb\endcsname \put(146,2410){\makebox(0,0)[r]{\strut{}}}\csname LTb\endcsname \put(146,2697){\makebox(0,0)[r]{\strut{}}}\csname LTb\endcsname \put(146,2984){\makebox(0,0)[r]{\strut{}}}\csname LTb\endcsname \put(146,3271){\makebox(0,0)[r]{\strut{}}}\csname LTb\endcsname \put(146,3558){\makebox(0,0)[r]{\strut{}}}\csname LTb\endcsname \put(473,1927){\makebox(0,0){\strut{}}}\csname LTb\endcsname \put(931,1927){\makebox(0,0){\strut{}}}\csname LTb\endcsname \put(1388,1927){\makebox(0,0){\strut{}}}\csname LTb\endcsname \put(1846,1927){\makebox(0,0){\strut{}}}\csname LTb\endcsname \put(2304,1927){\makebox(0,0){\strut{}}}}\gplgaddtomacro\gplfronttext{}\gplgaddtomacro\gplbacktext{\csname LTb\endcsname \put(2412,2123){\makebox(0,0)[r]{\strut{}}}\csname LTb\endcsname \put(2412,2410){\makebox(0,0)[r]{\strut{}}}\csname LTb\endcsname \put(2412,2697){\makebox(0,0)[r]{\strut{}}}\csname LTb\endcsname \put(2412,2984){\makebox(0,0)[r]{\strut{}}}\csname LTb\endcsname \put(2412,3271){\makebox(0,0)[r]{\strut{}}}\csname LTb\endcsname \put(2412,3558){\makebox(0,0)[r]{\strut{}}}\csname LTb\endcsname \put(2733,1927){\makebox(0,0){\strut{}}}\csname LTb\endcsname \put(3181,1927){\makebox(0,0){\strut{}}}\csname LTb\endcsname \put(3629,1927){\makebox(0,0){\strut{}}}\csname LTb\endcsname \put(4077,1927){\makebox(0,0){\strut{}}}\csname LTb\endcsname \put(4524,1927){\makebox(0,0){\strut{}}}}\gplgaddtomacro\gplfronttext{\csname LTb\endcsname \put(4280,3496){\makebox(0,0)[r]{\strut{}f150 Abs. Err}}\csname LTb\endcsname \put(4280,3300){\makebox(0,0)[r]{\strut{}f090$\times$f150 Rel. Err}}}\gplgaddtomacro\gplbacktext{\csname LTb\endcsname \put(146,629){\makebox(0,0)[r]{\strut{}}}\csname LTb\endcsname \put(146,1044){\makebox(0,0)[r]{\strut{}}}\csname LTb\endcsname \put(146,1459){\makebox(0,0)[r]{\strut{}}}\csname LTb\endcsname \put(146,1874){\makebox(0,0)[r]{\strut{}}}\csname LTb\endcsname \put(473,432){\makebox(0,0){\strut{}-0.2}}\csname LTb\endcsname \put(931,432){\makebox(0,0){\strut{}-0.1}}\csname LTb\endcsname \put(1388,432){\makebox(0,0){\strut{}0}}\csname LTb\endcsname \put(1846,432){\makebox(0,0){\strut{}0.1}}\csname LTb\endcsname \put(2304,432){\makebox(0,0){\strut{}0.2}}}\gplgaddtomacro\gplfronttext{\csname LTb\endcsname \put(2454,137){\makebox(0,0){\strut{}Recovered Phase (cycles)}}}\gplgaddtomacro\gplbacktext{\csname LTb\endcsname \put(2412,629){\makebox(0,0)[r]{\strut{}}}\csname LTb\endcsname \put(2412,1044){\makebox(0,0)[r]{\strut{}}}\csname LTb\endcsname \put(2412,1459){\makebox(0,0)[r]{\strut{}}}\csname LTb\endcsname \put(2412,1874){\makebox(0,0)[r]{\strut{}}}\csname LTb\endcsname \put(2733,432){\makebox(0,0){\strut{}-0.2}}\csname LTb\endcsname \put(3181,432){\makebox(0,0){\strut{}-0.1}}\csname LTb\endcsname \put(3629,432){\makebox(0,0){\strut{}0}}\csname LTb\endcsname \put(4077,432){\makebox(0,0){\strut{}0.1}}\csname LTb\endcsname \put(4524,432){\makebox(0,0){\strut{}0.2}}}\gplgaddtomacro\gplfronttext{\csname LTb\endcsname \put(4280,1946){\makebox(0,0)[r]{\strut{}f220 Abs. Err}}\csname LTb\endcsname \put(4280,1750){\makebox(0,0)[r]{\strut{}f090$\times$f220 Rel. Err}}\csname LTb\endcsname \put(4695,137){\makebox(0,0){\strut{} }}}\gplbacktext
    \put(0,0){\includegraphics[width={252.00bp},height={288.00bp}]{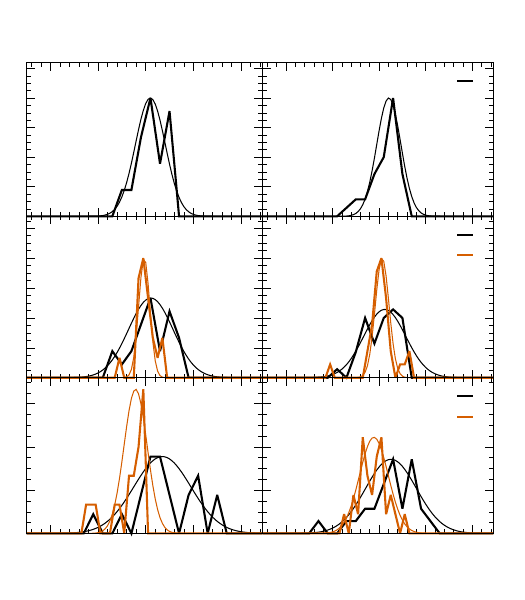}}\gplfronttext
  \end{picture}\endgroup
    \caption{Normalised histograms of recovered phases (thick lines), $t_0$, as well as phase differences, $t_{0,\nu_1} - t_{0,\nu_2}$. The overlaid Gaussian curves (thin lines) have widths derived from the median absolute deviation of the distribution, which serve as estimates of the absolute and relative phase errors. The errors are listed in Table~\ref{tab:act_phase_uncertainty}.}
   \label{plt:act_phase_uncertainty}
\end{figure}

\begin{table*}[tb]
  \centering
  \caption{Results of phase shift uncertainty analysis using sine-wave injections into ACT AGN light curves.}
  \begin{tabular}{|c|ccc|ccc|}
    \hline\hline
    AGN & Freq. & Count\tablefootmark{a} & Empirical & Freq. Pair & Empirical & MCMC \\
    & (GHz) & & Abs.\tablefootmark{b} (\%) & (GHz--GHz) & Rel. (\%) & (\%) \\
    \hline
             &  95 & 31 & 3.1 &  95--147 & 1.2 & 0.59 \\
    \pkstwo  & 147 & 25 & 4.7 &  95--225 & 2.4 & 1.8 \\
             & 225 & 21 & 6.0 & 147--225 & 3.9 & 1.8 \\
    \hline
             &  95 & 36 & 2.6 &  95--147 & 1.5 & 0.52 \\
    \pkszero & 147 & 37 & 4.5 &  95--225 & 3.1 & 1.5 \\
             & 225 & 28 & 5.5 & 147--225 & 3.3 & 1.5 \\
    \hline
  \end{tabular}
  \tablefoot{
  The MCMC error is obtained via the method described in Sec.~\ref{sec:sinfit_method}, and the empirical absolute and empirical relative errors are estimated by injecting known, artificial sine waves into real AGN light curves and measuring how well their phases can be recovered with respect to the input phase (absolute error) and with respect to the best fit in other frequencies (relative error). Note that it is not appropriate to compare the MCMC error directly against the absolute error; see text for details.\\
    \tablefoottext{a}{Number of AGN used (see text).}
    \tablefoottext{b}{All errors are quoted as percentage of a cycle, i.e., $100\times\Delta P / P_0$.}
  }
  \label{tab:act_phase_uncertainty}
\end{table*}

For our ACT light curves, we can make progress on the empirical front using our preliminary catalog of 205 bright AGN multi-frequency light curves (see Sec.~\ref{sec:actobs}). Under the assumption that the Type-A and Type-B variations affect blazars that are not SMBHB candidates in the same way that they affect the two SMBHB candidates discovered thus far -- which we stress is untested --  we can estimate the absolute and relative phase shift uncertainties by injecting a sinusoid into the light curves:
\begin{equation}\label{eq:sine_injection}
  f'(t) = f(t) + \snrsin\,\sigma_f\,\sin\left(\frac{2\pi t}{P_0}\right),
\end{equation}
where $f$ is the original light curve, $\sigma_f$ is the standard deviation of the original light curve, $\snrsin$ is the signal-to-noise ratio (SNR) of the sinusoid and $P_0$ its period. When we have multiple measurements at the same MJD at the same frequency (i.e., from different arrays), we use their mean value. Note that we have factored the amplitude of the sinusoid into a SNR term ($\snrsin$) and a noise term ($\sigma_f$) in order to use sine waves of similar SNR as our SMBHB candidates. Table~\ref{tab:smbhb_props} shows the SNR of the sinusoids in our SMBHB candidates at each frequency, which we obtained by subtracting the best-fitting sinusoid from the SMBHB candidate, measuring the standard deviation of its residuals, $\sigma_{\mathrm{resid}}$, and calculating:
\begin{equation}\label{eq:snrsin}
  \snrsin = \frac{A}{\sigma_{\mathrm{resid}}}.
\end{equation}
where $A$ is the best-fitting amplitude of the sine wave. In the same table, we also list the SNR of the residuals after subtracting the sine wave with respect to the flux uncertainties:
\begin{equation}\label{eq:snrvar}
  \snrvar = \frac{\sigma_{\mathrm{resid}}}{\widebar{\sigma}_{\mathrm{f}}}
\end{equation}
where $\widebar{\sigma}_{\mathrm{f}}$ is the mean value of the flux uncertainties in the light curve. This variable captures the SNR of the non-sinusoidal residuals.

We estimate the uncertainty in phase measurements by fitting a sinusoid to $f'(t)$,
\begin{equation}\label{eq:sin_fit}
  s(t) = A\sin\left[\frac{2\pi(t - t_0)}{P_0}\right] + b,
\end{equation}
where $A$, $t_0$ and $b$ are free parameters. A non-zero best-fitting $t_0$ represents an error in recovering the input phase. We carry out this fit on our ensemble light curves (with the SMBHBs excluded from the sample), where the injected light curves have $P_0$ and $\snrsin$ set to the values in Table~\ref{tab:smbhb_props}. We make two cuts on the ensemble of light curves. First, we require that the time span covered by each light curve be at least 95\% that of \pkstwo or \pkszero (see Table~\ref{tab:smbhb_props}) so that the length of the sine wave being fitted is representative of our SMBHB candidates. Second, we only include light curves which have $\snrvar < 5$, where here, $\snrvar = \sigma / \widebar{\sigma}_{\mathrm{f}}$, with $\sigma$ being the standard deviation of the original light curve. We explain the rationale for this cut in Sec.~\ref{sec:snr_considerations}, below. After both cuts, 10 to 18\% of the original 204 light curves remain, depending on the frequency and the AGN being studied (see `Count' in Table~\ref{tab:act_phase_uncertainty}).

With the final ensemble in hand, we use the \texttt{curve\_fit()} routine in the optimization package of \textsc{SciPy} to do a non-linear least squares fit to Eq.~\ref{eq:sin_fit} for each light curve and obtain a distribution of $t_0$ at each of the ACT frequencies. The width of this distribution indicates the absolute error on the phase shift. The width of the distribution of the difference of the recovered fits in the light curves at two frequencies, $\Delta t_{\nu_1,\nu_2} = t_{0,\nu_1} - t_{0,\nu_2}$, indicates the relative error on the phase shifts between frequencies $\nu_1$ and $\nu_2$.

\subsection{Results and discussion}

The distributions of our phase fits are shown in Fig.~\ref{plt:act_phase_uncertainty}. We estimate the width of the distribution using the median absolute deviation (MAD) statistic, since it is robust to outliers, and report errors as $\sigma = \mathrm{MAD}/0.6745$, where the numeric factor is the conversion to the Gaussian width for a normal distribution.\footnote{The MAD is the distance between the 50th and 75th percentiles (assuming the distribution is symmetric), which for a Gaussian is equal to $\sqrt{2}\,\mathrm{erf}^{-1}\!(\frac{1}{2})\sigma \approx 0.6745\sigma.$} These error estimates are listed in Table~\ref{tab:act_phase_uncertainty}. There is broad agreement between the scenarios representing the two SMBHB candidates: the absolute uncertainty is about 3\%, 5\% and 6\% of a cycle for 95, 147 and 225\,GHz, respectively, and the relative uncertainty between 95 and 147\,GHz is roughly 1.5\% and is about 2--4\% between 95/147 and 225\,GHz.

\begin{figure*}[tb]
   \centering
   \footnotesize
   \begingroup
  \makeatletter
  \providecommand\color[2][]{\GenericError{(gnuplot) \space\space\space\@spaces}{Package color not loaded in conjunction with
      terminal option `colourtext'}{See the gnuplot documentation for explanation.}{Either use 'blacktext' in gnuplot or load the package
      color.sty in LaTeX.}\renewcommand\color[2][]{}}\providecommand\includegraphics[2][]{\GenericError{(gnuplot) \space\space\space\@spaces}{Package graphicx or graphics not loaded}{See the gnuplot documentation for explanation.}{The gnuplot epslatex terminal needs graphicx.sty or graphics.sty.}\renewcommand\includegraphics[2][]{}}\providecommand\rotatebox[2]{#2}\@ifundefined{ifGPcolor}{\newif\ifGPcolor
    \GPcolortrue
  }{}\@ifundefined{ifGPblacktext}{\newif\ifGPblacktext
    \GPblacktexttrue
  }{}\let\gplgaddtomacro\g@addto@macro
\gdef\gplbacktext{}\gdef\gplfronttext{}\makeatother
  \ifGPblacktext
\def\colorrgb#1{}\def\colorgray#1{}\else
\ifGPcolor
      \def\colorrgb#1{\color[rgb]{#1}}\def\colorgray#1{\color[gray]{#1}}\expandafter\def\csname LTw\endcsname{\color{white}}\expandafter\def\csname LTb\endcsname{\color{black}}\expandafter\def\csname LTa\endcsname{\color{black}}\expandafter\def\csname LT0\endcsname{\color[rgb]{1,0,0}}\expandafter\def\csname LT1\endcsname{\color[rgb]{0,1,0}}\expandafter\def\csname LT2\endcsname{\color[rgb]{0,0,1}}\expandafter\def\csname LT3\endcsname{\color[rgb]{1,0,1}}\expandafter\def\csname LT4\endcsname{\color[rgb]{0,1,1}}\expandafter\def\csname LT5\endcsname{\color[rgb]{1,1,0}}\expandafter\def\csname LT6\endcsname{\color[rgb]{0,0,0}}\expandafter\def\csname LT7\endcsname{\color[rgb]{1,0.3,0}}\expandafter\def\csname LT8\endcsname{\color[rgb]{0.5,0.5,0.5}}\else
\def\colorrgb#1{\color{black}}\def\colorgray#1{\color[gray]{#1}}\expandafter\def\csname LTw\endcsname{\color{white}}\expandafter\def\csname LTb\endcsname{\color{black}}\expandafter\def\csname LTa\endcsname{\color{black}}\expandafter\def\csname LT0\endcsname{\color{black}}\expandafter\def\csname LT1\endcsname{\color{black}}\expandafter\def\csname LT2\endcsname{\color{black}}\expandafter\def\csname LT3\endcsname{\color{black}}\expandafter\def\csname LT4\endcsname{\color{black}}\expandafter\def\csname LT5\endcsname{\color{black}}\expandafter\def\csname LT6\endcsname{\color{black}}\expandafter\def\csname LT7\endcsname{\color{black}}\expandafter\def\csname LT8\endcsname{\color{black}}\fi
  \fi
    \setlength{\unitlength}{0.0500bp}\ifx\gptboxheight\undefined \newlength{\gptboxheight}\newlength{\gptboxwidth}\newsavebox{\gptboxtext}\fi \setlength{\fboxrule}{0.5pt}\setlength{\fboxsep}{1pt}\definecolor{tbcol}{rgb}{1,1,1}\begin{picture}(10080.00,2580.00)\gplgaddtomacro\gplbacktext{\csname LTb\endcsname \put(441,393){\makebox(0,0)[r]{\strut{}$0$}}\csname LTb\endcsname \put(441,885){\makebox(0,0)[r]{\strut{}$2$}}\csname LTb\endcsname \put(441,1378){\makebox(0,0)[r]{\strut{}$4$}}\csname LTb\endcsname \put(441,1870){\makebox(0,0)[r]{\strut{}$6$}}\csname LTb\endcsname \put(441,2363){\makebox(0,0)[r]{\strut{}$8$}}\csname LTb\endcsname \put(2624,196){\makebox(0,0){\strut{}f090--f150}}\csname LTb\endcsname \put(3667,196){\makebox(0,0){\strut{}f090--f220}}\csname LTb\endcsname \put(4709,196){\makebox(0,0){\strut{}f150--f220}}}\gplgaddtomacro\gplfronttext{\csname LTb\endcsname \put(1009,2186){\makebox(0,0)[l]{\strut{}$\snrsin$ of PKS 2131 (3.4, 2.5, 1.1)}}\csname LTb\endcsname \put(1009,1989){\makebox(0,0)[l]{\strut{}$\snrsin = 0$}}\csname LTb\endcsname \put(1009,1793){\makebox(0,0)[l]{\strut{}$\snrsin = 1$}}\csname LTb\endcsname \put(1009,1596){\makebox(0,0)[l]{\strut{}$\snrsin = 2$}}\csname LTb\endcsname \put(1009,1400){\makebox(0,0)[l]{\strut{}$\snrsin = 3$}}\csname LTb\endcsname \put(1009,1203){\makebox(0,0)[l]{\strut{}$\snrsin = 5$}}\csname LTb\endcsname \put(171,1378){\rotatebox{-270.00}{\makebox(0,0){\strut{}Rel. Uncertainty (\% cycle)}}}}\gplgaddtomacro\gplbacktext{\csname LTb\endcsname \put(5133,393){\makebox(0,0)[r]{\strut{}}}\csname LTb\endcsname \put(5133,885){\makebox(0,0)[r]{\strut{}}}\csname LTb\endcsname \put(5133,1378){\makebox(0,0)[r]{\strut{}}}\csname LTb\endcsname \put(5133,1870){\makebox(0,0)[r]{\strut{}}}\csname LTb\endcsname \put(5133,2363){\makebox(0,0)[r]{\strut{}}}\csname LTb\endcsname \put(7246,196){\makebox(0,0){\strut{}f090--f150}}\csname LTb\endcsname \put(8254,196){\makebox(0,0){\strut{}f090--f220}}\csname LTb\endcsname \put(9262,196){\makebox(0,0){\strut{}f150--f220}}}\gplgaddtomacro\gplfronttext{\csname LTb\endcsname \put(5701,2186){\makebox(0,0)[l]{\strut{}All $\snrvar$}}\csname LTb\endcsname \put(5701,1989){\makebox(0,0)[l]{\strut{}$\snrvar < 3$}}\csname LTb\endcsname \put(5701,1793){\makebox(0,0)[l]{\strut{}$\snrvar < 4$}}\csname LTb\endcsname \put(5701,1596){\makebox(0,0)[l]{\strut{}$\snrvar < 5$}}\csname LTb\endcsname \put(5701,1400){\makebox(0,0)[l]{\strut{}$\snrvar < 5$, $\snrsin = 3$}}\csname LTb\endcsname \put(5701,1203){\makebox(0,0)[l]{\strut{}$\snrvar < 6$}}\csname LTb\endcsname \put(5059,1378){\rotatebox{-270.00}{\makebox(0,0){\strut{}}}}}\gplbacktext
    \put(0,0){\includegraphics[width={504.00bp},height={129.00bp}]{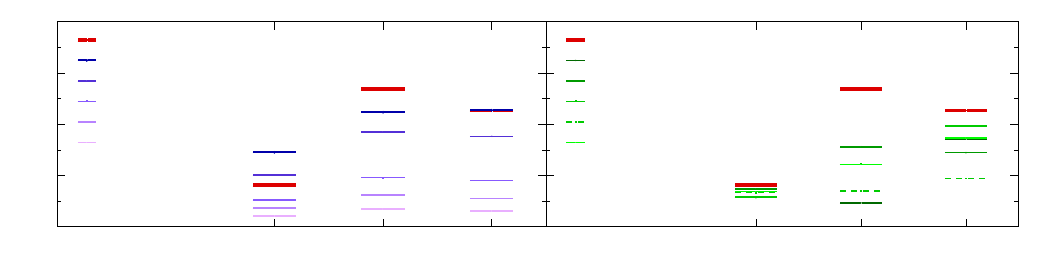}}\gplfronttext
  \end{picture}\endgroup
    \caption{Relative errors calculated subject to various SNR values used in the sine waves injected into our ensemble of AGN light curves. \textit{Left:} In this panel, we compare the relative errors obtained when $\snrsin$ is set to the value from \pkstwo (heavy red lines), to when $\snrsin$ is set, at all frequencies, to 0, 1, 2, 3 or 5 (blue lines). \textit{Right:} Relative errors obtained when AGN are cut from the sample based on their $\snrvar$ at 95\,GHz (green lines), compared to what is obtained when all AGN are used (red lines). The dashed green line shows results when both an $\snrvar < 5$ cut is applied and $\snrsin$ is set to 3 at all frequencies.}
   \label{plt:act_phase_snr}
\end{figure*}

The uncertainties from the MCMC fits (Sec.~\ref{sec:sinfit_method}) are also shown in Table~\ref{tab:act_phase_uncertainty}. Here, we have assumed the hypothesis that the SMBHB light curves are dominated by Type-A variations, such that the MCMC uncertainties represent purely \textit{relative errors}, and have added in quadrature the uncertainties in $\phi_0$ provided in Tables~\ref{tab:2131_fits} and \ref{tab:0805_fits} for the three pairs of ACT frequencies. The MCMC uncertainties are ${\sim}$2 times smaller than our empirical estimates. Given our current limited understanding of SMBHB light curve properties, we consider MCMC and empirical relative uncertainties to be in broad agreement. The MCMC uncertainties could be artificially low if Type-B variations are non-negligble. On the other hand, the empirical analysis is based on the assumption that injecting sine waves into non-SMBHB AGN light curves provides an accurate simulation of SMBHB light curves, or, in other words, that Type-A and Type-B variations behave the same way in SMBHB and non-SMBHB AGN. If this assumption is faulty then the resulting errors could be inflated; furthermore, the sample size is rather small. For these reasons, the MCMC uncertainties may well be appropriate for SMBHBs, and that they appear to be reasonable given the fit of the simple quadratic curve in Fig.~\ref{plt:2131_phases}.  In the end, the main results of this paper are not sensitive to the exact size of the error bars, but the work in this Appendix demonstrates the need for more data to better understand the uncertainties in SMBHB sine fits.

\subsection{Further tests on phase uncertainty estimates}\label{sec:snr_considerations}

\paragraph{Varying $\snrsin$:} The left panel of Fig.~\ref{plt:act_phase_snr} shows the effect on the relative uncertainties of varying the value of $\snrsin$ injected into the ensemble of light curves. Of particular interest is that the 95--225\,GHz and 147--225\,GHz uncertainties based on the fiducial $\snrsin$ values, taken from our real SMBHB candidate (Table~\ref{tab:smbhb_props}), are essentially the same if \textit{no} sine wave is inserted into light curves in our ensemble, i.e., $\snrsin = 0$. This means that our errors relative to the 225\,GHz light curves, which only have $\snrsin \approx 1$, are dominated by Type-A variations: it is not so much the phase shift in the sine wave that is being measured as the shift in the Type-A variations. On the other hand, the 95--147\,GHz relative uncertainty, where the SMBHB candidate has $\snrsin \approx 2.5{-}3.5$, is significantly smaller than if $\snrsin$ is set to zero. When $\snrsin$ is raised to 3 for light curves at all frequencies, the relative errors between all frequencies is ${\approx}1\%$, and is reduced as $\snrsin$ increases, as expected.

\paragraph{Varying $\snrvar$:} The right panel of Fig.~\ref{plt:act_phase_snr} explores how well the relative uncertainty can be measured for different $\snrvar$. Our SMBHBs have comparatively small $\snrvar$: at 95\,GHz, $\snrvar = 3.8$ and 2.0 for \pkstwo and \pkszero, respectively, compared to the median of 9.4 for the 204 other AGN in the ensemble. Effectively, this means that the residuals of the SMBHB candidates exhibit lower variations than are found in the other, non-SMBHB light curves. As the figure shows, when all AGN are used to estimate the relative light curves, the errors relative to 225\,GHz are higher than if we perform a cut on $\snrvar$, such that the AGN included in the sample have variations that are more representative of the SMBHB candidates. On the other hand, the 95--147\,GHz relative error is not sensitive to any cuts on $\snrvar$. Again, the interpretation is that when $\snrsin \gtrsim 2$, the sine wave is larger than the other variations, and thus the relative phase errors are not very sensitive to the non-sinusoidal variations; this is not the case when $\snrsin \sim 1$. With our limited sample, doing a cut for $\snrvar < 3$ at 95\,GHz, which would best represent our SMBHB candidates, only leaves 13 AGN in the sample. We therefore opt for a cut on $\snrvar < 5$, which more than doubles the number of available AGN (see the `Count' column in Table~\ref{tab:act_phase_uncertainty}) while not significantly altering the resulting error estimates, as can be seen in Fig.~\ref{plt:act_phase_snr}. Finally, as a consistency check, for our fiducial cut of $\snrvar < 5$ we also did a test where we set $\snrsin=3$ for all frequencies, and found results similar to the left-hand panel of Fig.~\ref{plt:act_phase_snr}, where no cut on $\snrvar$ was used. This confirms the point made above, that when $\snrsin \gtrsim 2$, non-sinusoidal variations do not impact the uncertainty much.

\begin{figure}[tb]
   \centering
   \footnotesize
   \begingroup
  \makeatletter
  \providecommand\color[2][]{\GenericError{(gnuplot) \space\space\space\@spaces}{Package color not loaded in conjunction with
      terminal option `colourtext'}{See the gnuplot documentation for explanation.}{Either use 'blacktext' in gnuplot or load the package
      color.sty in LaTeX.}\renewcommand\color[2][]{}}\providecommand\includegraphics[2][]{\GenericError{(gnuplot) \space\space\space\@spaces}{Package graphicx or graphics not loaded}{See the gnuplot documentation for explanation.}{The gnuplot epslatex terminal needs graphicx.sty or graphics.sty.}\renewcommand\includegraphics[2][]{}}\providecommand\rotatebox[2]{#2}\@ifundefined{ifGPcolor}{\newif\ifGPcolor
    \GPcolortrue
  }{}\@ifundefined{ifGPblacktext}{\newif\ifGPblacktext
    \GPblacktexttrue
  }{}\let\gplgaddtomacro\g@addto@macro
\gdef\gplbacktext{}\gdef\gplfronttext{}\makeatother
  \ifGPblacktext
\def\colorrgb#1{}\def\colorgray#1{}\else
\ifGPcolor
      \def\colorrgb#1{\color[rgb]{#1}}\def\colorgray#1{\color[gray]{#1}}\expandafter\def\csname LTw\endcsname{\color{white}}\expandafter\def\csname LTb\endcsname{\color{black}}\expandafter\def\csname LTa\endcsname{\color{black}}\expandafter\def\csname LT0\endcsname{\color[rgb]{1,0,0}}\expandafter\def\csname LT1\endcsname{\color[rgb]{0,1,0}}\expandafter\def\csname LT2\endcsname{\color[rgb]{0,0,1}}\expandafter\def\csname LT3\endcsname{\color[rgb]{1,0,1}}\expandafter\def\csname LT4\endcsname{\color[rgb]{0,1,1}}\expandafter\def\csname LT5\endcsname{\color[rgb]{1,1,0}}\expandafter\def\csname LT6\endcsname{\color[rgb]{0,0,0}}\expandafter\def\csname LT7\endcsname{\color[rgb]{1,0.3,0}}\expandafter\def\csname LT8\endcsname{\color[rgb]{0.5,0.5,0.5}}\else
\def\colorrgb#1{\color{black}}\def\colorgray#1{\color[gray]{#1}}\expandafter\def\csname LTw\endcsname{\color{white}}\expandafter\def\csname LTb\endcsname{\color{black}}\expandafter\def\csname LTa\endcsname{\color{black}}\expandafter\def\csname LT0\endcsname{\color{black}}\expandafter\def\csname LT1\endcsname{\color{black}}\expandafter\def\csname LT2\endcsname{\color{black}}\expandafter\def\csname LT3\endcsname{\color{black}}\expandafter\def\csname LT4\endcsname{\color{black}}\expandafter\def\csname LT5\endcsname{\color{black}}\expandafter\def\csname LT6\endcsname{\color{black}}\expandafter\def\csname LT7\endcsname{\color{black}}\expandafter\def\csname LT8\endcsname{\color{black}}\fi
  \fi
    \setlength{\unitlength}{0.0500bp}\ifx\gptboxheight\undefined \newlength{\gptboxheight}\newlength{\gptboxwidth}\newsavebox{\gptboxtext}\fi \setlength{\fboxrule}{0.5pt}\setlength{\fboxsep}{1pt}\definecolor{tbcol}{rgb}{1,1,1}\begin{picture}(5040.00,2880.00)\gplgaddtomacro\gplbacktext{\csname LTb\endcsname \put(441,629){\makebox(0,0)[r]{\strut{}$0$}}\csname LTb\endcsname \put(441,1035){\makebox(0,0)[r]{\strut{}$1$}}\csname LTb\endcsname \put(441,1442){\makebox(0,0)[r]{\strut{}$2$}}\csname LTb\endcsname \put(441,1849){\makebox(0,0)[r]{\strut{}$3$}}\csname LTb\endcsname \put(441,2256){\makebox(0,0)[r]{\strut{}$4$}}\csname LTb\endcsname \put(441,2663){\makebox(0,0)[r]{\strut{}$5$}}\csname LTb\endcsname \put(785,432){\makebox(0,0){\strut{}$-15$}}\csname LTb\endcsname \put(1401,432){\makebox(0,0){\strut{}$-10$}}\csname LTb\endcsname \put(2017,432){\makebox(0,0){\strut{}$-5$}}\csname LTb\endcsname \put(2632,432){\makebox(0,0){\strut{}$0$}}\csname LTb\endcsname \put(3248,432){\makebox(0,0){\strut{}$5$}}\csname LTb\endcsname \put(3864,432){\makebox(0,0){\strut{}$10$}}\csname LTb\endcsname \put(4479,432){\makebox(0,0){\strut{}$15$}}}\gplgaddtomacro\gplfronttext{\csname LTb\endcsname \put(1303,2486){\makebox(0,0)[l]{\strut{}\pkstwo}}\csname LTb\endcsname \put(1303,2289){\makebox(0,0)[l]{\strut{}\pkszero}}\csname LTb\endcsname \put(171,1646){\rotatebox{-270.00}{\makebox(0,0){\strut{}Relative Error (\% cycle)}}}\csname LTb\endcsname \put(2632,137){\makebox(0,0){\strut{}Input Phase Shift (\% cycle)}}}\gplbacktext
    \put(0,0){\includegraphics[width={252.00bp},height={144.00bp}]{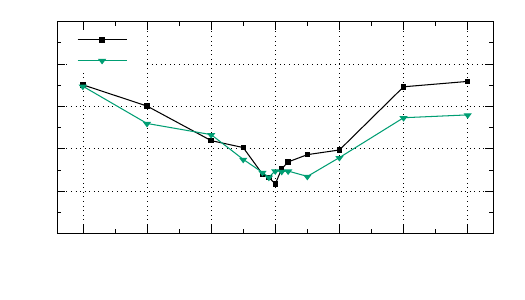}}\gplfronttext
  \end{picture}\endgroup
    \caption{The effect of adding relative phase shifts into the injected sinusoids. Shown here is the 95--147\,GHz relative error, where the sine wave injected into the 147\,GHz light curve has been shifted relative to the 95\,GHz light curve by the amount shown on the $x$-axis. Results for sine waves having properties of each of \pkstwo and \pkszero (see Table~\ref{tab:smbhb_props}) are shown.}
   \label{plt:act_sim-shifts}
\end{figure}

\paragraph{Including phase shifts in injected sine waves:} An inaccuracy in our empirical error estimates is that we inject sine waves with the same phase at all frequencies. In reality, we know from our two SMBHB candidates that they are slightly phase shifted. To test whether neglecting this effect affects our error estimates, we introduced different relative phase shifts into the injected sine waves and repeated the analysis. As for our fiducial error estimates, we apply a cut of $\snrvar < 5$. Fig.~\ref{plt:act_sim-shifts} shows the results for the 95--147\,GHz relative uncertainty, and demonstrates that it does have an effect on the error estimate. However, because the measured phase shifts between the ACT frequencies in our SMBHB candidates are smaller than 2\%, corresponding to the central points along the $x$-axis in the figure, neglecting this effect in our fiducial results (Table~\ref{tab:act_phase_uncertainty}) only makes a modest difference. The results for the 95--225\,GHz and 147--225\,GHz uncertainties are considerably noisier but broadly similar.\footnote{The 95--225\,GHz result for \pkstwo is the noisiest, with the uncertainty estimate varying by a factor of 2.5 for shifts between $-2\%$ and $2\%$.} In future estimates of the phase shift uncertainties, this effect should be taken into consideration.

\begin{table*}[htb]
\centering
\caption{Generalized Lomb-Scargle test results of the optical light curves of \pkstwo.}
\label{tab:periodicityanalysis}
\begin{tabular}{lllrrr}
    \hline \hline
    Test & Description & Data & GLS period (days) & GLS power & p-value \\
    \hline
    1 & Independent of radio data & CTRS+ZTF & 1711 & 0.22 & 0.937 \\
    2 & Prior knowledge of radio period & CTRS+ZTF & & & 0.047 \\
    3 & Prior knowledge of radio period & ZTF & & & 0.065 \\
    4 & Prior knowledge of radio period & CRTS & & & 0.079 \\
    \hline
\end{tabular}
\tablefoot{The  GLS peak in the OVRO 15~GHz light curve has power $\mathcal{P}_{\rm peak}\,=\,0.83$, and p-value $=5.3 \times 10^{-5}$ (K25).}
\end{table*}

\section{Optical light curve periodicity}
\label{app:sim}

We analyzed the joint CRTS V-band and ZTF g-band light curve of \pkstwo with the generalized Lomb-Scargle (GLS) periodogram \citep{1976Ap&SS..39..447L, 1982ApJ...263..835S, 2009A&A...496..577Z}. The strongest peak in the periodogram is at a period of 1711 days and a power of 0.22 (normalized between 0 and 1, where 1 would imply a perfect sine fit). The period is within the 3$\sigma$ range, 1657.5--1854.1~days, of the periodicity detected in the OVRO 15\,GHz data as reported in Table~1 of K25. The low GLS peak power is a result of the strong red-noise component in the data.  We estimate\footnote{We used this Python package for the power spectral density analysis: \url{https://github.com/skiehl/psd_analysis}.} a power-law index of $\beta_\mathrm{opt} = 1.5$ of the red-noise power spectral density (PSD $\sim \nu^{-\beta}$, where $\nu$ is the temporal frequency). Compared to the 15\,GHz radio light curve with a PSD index $\beta_\mathrm{radio} = 1.8$ (O22), the optical PSD is flatter, implying stronger variability on short time scales.

GLS analysis of the joint CRTS and ZTF g-band data gives a period of 1711 days and a power of 0.22. ZTF alone gives a period of 1728 days and a power of 0.23 whereas CRTS alone does not identify a period consistent with the radio period. In fact, the strongest peaks in the CRTS periodogram come from the sampling window associated with the CRTS observing cadence. The effect of the CRTS cadence can be demonstrated by downsampling the ZTF data (median $\Delta t = 3$\,days) to the CRTS cadence (median $\Delta  t = 15$\,days): a joint analysis gives a period of 1724 days and a power of 0.36 but the false alarm probability (FAP) from bootstrap analysis shows the result is no longer significant. Although the FAP is not relevant for assessing the statistical significance of any peak against a correlated noise background, it can be used to assess the effect of cadence on a signal. In this case, the shorter cadence of ZTF data improves the statistical detectability of a signal in the joint data to a significant level, with a p-value of 0.002, as compared to a p-value of 0.414 for joint data with the same median cadence.

We perform different four tests to evaluate the feasibility of using optical data to robustly detect sinusoidal variations in \pkstwo.

\paragraph{Test 1:} We use the procedure described in Appendix~A of O22 to estimate the significance of the detected periodicity. Our null hypothesis is that the detected periodicity is a spurious result of a pure red-noise stochastic process. We simulate 20\,000 artificial light curves that have the same PSD, probability distribution function, time sampling and observational noise characteristics as the optical light curve. We use the \verb|lcsim| python package \citep{2023ascl.soft10002K} based on the algorithm of \citet{2013MNRAS.433..907E}. For the original data and each simulation we calculate the GLS periodogram, identify the strongest peak, and use the remaining simulations to estimate the probability that the red-noise process produces a power at least as strong as detected at the peak frequency. This probability is used as test statistic in the next analysis step. Among all simulations we count those that have a test statistic at least as low as that derived from the original light curve, which gives us the \textit{global p-value}. This global p-value takes into account that spurious periodic signals may occur at any frequency in the frequency range tested by the GLS periodogram. We refer to Appendices~A in O22 and D25 for a detailed description of the procedure.
We find a global p-value of 0.937. Therefore, we cannot reject the null hypothesis of a red-noise process producing a spurious periodicity. 

\paragraph{Test 2:} When we take our prior knowledge of the strongly significant radio periodicity (p-value $= 5.3 \times 10^{-7}$; K25) into account, we can ask the following question: How likely is it that the red-noise process produces a GLS peak at least as strong as observed in the optical data with a period in the $3\sigma$ range of the radio period (1657.5--1854.1\,days, K25)? Among the 20\,000 simulations we count 934 that fulfill these criteria, corresponding to a p-value of 0.047. Even with the optical period falling close to the radio period, we cannot confidently rule out the red-noise hypothesis.

Stronger support for the true existence of a period in the optical data consistent with the period seen in the radio data comes from the fact of how well the phase of the optical periodicity ties into the phase-frequency relation discussed in K25.

\paragraph{Tests~3 and~4:} We also analysed the CRTS and ZTF data independently. The GLS periodogram of the ZTF data shows the strongest peak at a period of 1728~days. The periodogram of the CRTS also shows a peak near the radio periodicity. However, there are several other spurious peaks that do not correspond to significant detections. Using the same approach as for Test~2 above on the individual light curves, in Table~\ref{tab:periodicityanalysis} we see that the combined CRTS+ZTF light curve provides the most evidence against the null hypothesis. This is expected as the joint data cover more cycles of the long time-scale periodicity, hidden in the red-noise component, that produces strong variability on shorter time-scales.

The foregoing analyses demonstrates that it can be more complicated to detect a
periodicity in optical blazar light curves from current surveys,
since the short-term variability is stronger than at radio-submm
frequencies. Thus more extended long-term observations with higher
cadence are needed to reveal periodicities in the presence of the
red-noise component

\end{appendix}

\end{document}